\documentclass[prd,twocolumn,eqsecnum,nofootinbib]{revtex4}
\usepackage{bm} 
\usepackage{amssymb}
\usepackage{graphicx} 
\newcommand{\base}[2]{e^{#1}_{#2}}
\begin{document}

\title{Retarded coordinates based at a world line, and the motion of a 
small black hole in an external universe}
\author{Eric Poisson}
\affiliation{Department of Physics, University of Guelph, Guelph, 
Ontario, Canada N1G 2W1}
\affiliation{Perimeter Institute for Theoretical Physics,  
35 King Street North, Waterloo, Ontario, Canada N2J 2W9} 
\date{October 24, 2003} 
\begin{abstract} 
In the first part of this article I present a system of retarded
coordinates based at an arbitrary world line of an arbitrary curved 
spacetime. The retarded-time coordinate labels forward light cones
that are centered on the world line, the radial coordinate is an affine
parameter on the null generators of these light cones, and the angular 
coordinates are constant on each of these generators. The spacetime
metric in the retarded coordinates is displayed as an expansion in
powers of the radial coordinate and expressed in terms of the world
line's acceleration vector and the spacetime's Riemann tensor
evaluated at the world line. The formalism is illustrated in two
examples, the first involving a comoving world line of a
spatially-flat cosmology, the other featuring an observer in circular
motion in the Schwarzschild spacetime. The main application of the
formalism is presented in the second part of the article, in which I
consider the motion of a small black hole in an empty external
universe. I use the retarded coordinates to construct the metric of
the small black hole perturbed by the tidal field of the external
universe, and the metric of the external universe perturbed by the
presence of the black hole. Matching these metrics produces the
MiSaTaQuWa equations of motion for the small black hole. 
\end{abstract} 
\maketitle

\section{Introduction} 

In the first part of this article I present a system of retarded 
coordinates $(u,r,\theta,\phi)$ based at an arbitrary world line
$\gamma$ of an arbitrary spacetime with metric $g_{\alpha\beta}$. The 
coordinates are adapted to the forward light cone of each point
$z(\tau)$ of the world line ($\tau$ is the proper-time parameter on
$\gamma$). The retarded-time coordinate $u$ is constant on each light
cone, and it agrees with $\tau$ at the cone's apex. The radial
coordinate $r$ is an affine parameter on the cone's null generators,
and it gives a measure of distance away from the world line. The
angular coordinates $\theta^A = (\theta,\phi)$ are constant on each of
these generators. 

This formalism complements a line of research that was initiated by
Synge \cite{synge:60} and pursued by Ellis and his collaborators 
\cite{ellis-etal:85, ellis-etal:92a, ellis-etal:92b, ellis-etal:92c,
ellis-etal:92d, ellis-etal:94} in their work on observational
cosmology. While the central ideas exploited here are the same as with 
Synge and Ellis, my implementation is substantially different: While
Synge and Ellis sought definitions for their optical or observational 
coordinates that apply to large regions of the spacetime, my
considerations are limited to a small neighborhood of the world line;
and while Ellis favored a construction based on past light cones
(because of the cosmological context in which information at $\gamma$
is gathered from the past), here the preference is given to future
light cones.  

The introduction of retarded coordinates is motivated by the desire to
construct solutions of wave equations for massless fields that are
produced by pointlike sources moving on the world line. The retarded
coordinates naturally incorporate the causal relation that exists
between the source and the field, and for this reason the solution
takes a simple explicit form (in the neighborhood in which the
coordinates are defined). The point of view developed here, therefore,
is opposite to the cosmological view developed by Ellis: here $\gamma$
is the cause and the effect is propagated along forward light cones. 

A quasi-Cartesian version of the retarded coordinates is introduced in 
Sec.~II B, after reviewing some necessary geometrical elements (a 
tetrad transported on $\gamma$, Synge's world function, and the
bitensor of parallel transport) in Sec.~II A. In Sec.~II C, I explore
the significance of $r$ as an affine parameter on the generators of 
the light cones, and describe the vector $k^\alpha$ that is tangent to   
the congruence of generators. The metric in retarded coordinates is
gradually constructed in Secs.~II D through F, and its quasi-Cartesian
form is displayed in Eqs.~(\ref{2.6.3})--(\ref{2.6.5}). In Sec.~II G,
I carry out a transformation to angular coordinates, and this form of 
the metric is displayed in Eqs.~(\ref{2.7.8})--(\ref{2.7.11}). In
Sec.~II H, I present an important simplification of the formalism that 
occurs when $\gamma$ is a geodesic and the Ricci tensor vanishes on
the world line. Two examples are worked out in Secs.~II I and J: first I
apply the formalism to a comoving world line of a spatially-flat
cosmology, and then I examine the metric near an observer in circular
motion in the Schwarzschild spacetime.  

The main application of the formalism is presented in the 
second part of the paper. Here I consider the motion of a nonrotating
black hole of small mass $m$ in a background spacetime with metric
$g_{\alpha\beta}$; the metric is assumed to be a solution to the
Einstein field equations in vacuum. By employing the powerful method
of matched asymptotic expansions (see, for example,
Ref.~\cite{thorne-hartle:85} for a discussion), I derive equations of
motion for the black hole. If $z^\mu(\tau)$ are the parametric
relations describing the black hole's world line, $u^\mu = 
d z^\mu/d\tau$ is the velocity vector, and $a^\mu = D u^\mu/d\tau$ the
acceleration vector, then the equations of motion take the form of 
\begin{equation}
a^\mu = -\frac{1}{2} \bigl( g^{\mu\nu} + u^\mu
u^\nu \bigr) \bigl( 2 h^{\rm tail}_{\nu\lambda\rho} 
- h^{\rm tail}_{\lambda\rho\nu} \bigr) u^\lambda u^\rho,  
\label{1.1} 
\end{equation}   
where  
\begin{eqnarray}
h^{\rm tail}_{\mu\nu\lambda} &=& 4 m \int_{-\infty}^{\tau^-} 
\nabla_\lambda \biggl( G_{\mu\nu\mu'\nu'}
- \frac{1}{2} g_{\mu\nu} G^{\rho}_{\ \rho\mu'\nu'}
\biggr)(\tau,\tau') 
\nonumber \\ & & \hspace*{40pt} \mbox{} \times 
u^{\mu'} u^{\nu'}\, d\tau' 
\label{1.2} 
\end{eqnarray}     
is an integration over the past portion of the world line; the
integral involves the retarded Green's function 
$G_{\mu\nu\mu'\nu'}(z(\tau),z(\tau'))$ for the metric perturbation
\cite{sciama-etal:69} associated with the small black hole, and it is
cut off at $\tau' = \tau^- \equiv \tau - 0^+$ to avoid the singular
behavior of the Green's function at coincidence.      

Equation (\ref{1.1}) is not new, and the method of derivation
presented here also is not new; they both originate in a 1997 paper by  
Mino, Sasaki, and Tanaka \cite{mino-etal:97}. (The equations were
later rederived by Quinn and Wald \cite{quinn-wald:97}, and they are
now known as the MiSaTaQuWa equations of motion.) The method of
matched asymptotic expansions was first used to establish that in the
limit $m \to 0$, the motion of a black hole is geodesic in the
background spacetime \cite{manasse:63, kates:80, death:96}. Using this
approach, Mino, Sasaki, and Tanaka (see also Ref.~\cite{detweiler:01})
were able to compute the first-order correction to this result, and
showed that at order $m$, the motion is accelerated and governed by
Eq.~(\ref{1.1}). Their derivation, however, did not rely on a specific
choice of coordinates, and it left many details obscured. By adopting 
retarded coordinates, these details can be revealed and the matching
procedure clarified. This was my aim here: to provide a clearer, more
explicit implementation of the matching procedure employed by Mino, 
Sasaki, and Tanaka.                

The method of matched asymptotic expansions is explained in Sec.~III 
A. Essentially, the metric of the small black hole perturbed by the 
tidal gravitational field of the external universe is matched to the
metric of the external universe perturbed by the black hole; ensuring
that this metric is a valid solution to the vacuum Einstein field
equations determines the motion of the black hole. The perturbed
metric of the small black hole is constructed in an {\it internal
zone} in terms of internal retarded coordinates
$(\bar{u},\bar{t},\bar{\theta}^A)$; this 
is carried out in Sec.~III B. The perturbed metric of the external
universe is constructed in an {\it external zone} in terms of external
retarded coordinates $(u,r,\theta^A)$; this is carried out in Sec.~III
C. The matching of the two metrics is performed in a {\it buffer zone}
that overlaps with both the internal and external zones; this involves
a transformation from the external to the internal coordinates that is
explicitly worked out in Appendix C. The matching is carried out in
Sec.~III D, where we see that it does indeed produce Eq.~(\ref{1.1}).   

This concludes the summary of the results contained in this
paper. Various technical details are relegated to Appendices A and 
B. Throughout the paper I work in geometrized units ($G=c=1$) and with
the conventions of Misner, Thorne, and Wheeler \cite{MTW:73}. Other
applications of the retarded coordinates (including the fields
produced by point particles, and their motion in curved spacetime) are
featured in a recent review article \cite{poisson:03}.      

\section{Retarded coordinates} 

\subsection{Geometrical elements}

To construct the retarded coordinates we must first introduce some
geometrical elements on the world line $\gamma$ at which the
coordinates are based. The world line is described by relations
$z^\mu(\tau)$ with $\tau$ denoting proper time, its normalized tangent
vector is $u^\mu = dz^\mu/d\tau$, and its acceleration vector is 
$a^\mu = D u^\mu/d\tau$ with $D/d\tau$ denoting covariant
differentiation along the world line. Throughout we use
Greek indices $\mu$, $\nu$, $\lambda$, $\rho$, etc.\ to refer to
tensor fields defined, or evaluated, on the world line.    

We install on $\gamma$ an orthonormal tetrad that consists of the 
tangent vector $u^\mu$ and three spatial vectors $\base{\mu}{a}$. 
These are transported on the world line according to    
\begin{equation}
\frac{D \base{\mu}{a}}{d \tau} = a_a u^{\mu} 
+ \omega_a^{\ b} \base{\mu}{b},  
\label{2.1.1}
\end{equation}  
where $a_a(\tau) = a_{\mu} \base{\mu}{a}$ are the frame components of 
the acceleration vector and $\omega_{ab}(\tau) = -\omega_{ba}(\tau)$
is a prescribed rotation tensor. Setting $\omega_{ab} = 0$ would make
the triad Fermi-Walker transported on the world line, and in many
applications this would be a sensible choice. We nevertheless allow
the spatial vectors to rotate as they are transported on the world
line; this makes the formalism more general, and we will need
this flexibility in Sec.~III of the paper. It is easy to check that
Eq.~(\ref{2.1.1}) is compatible with the requirement that the tetrad 
$(u^\mu,\base{\mu}{a})$ be orthonormal everywhere on
$\gamma$. 

From the tetrad on $\gamma$ we define a dual tetrad
$(\base{0}{\mu},\base{a}{\mu})$ with the relations 
\begin{equation} 
\base{0}{\mu} = -u_\mu, \qquad 
\base{a}{\mu} = \delta^{ab} g_{\mu\nu} \base{\nu}{b}. 
\label{2.1.2} 
\end{equation} 
The dual vectors $\base{a}{\mu}$ satisfy a transport law that is very
similar to Eq.~(\ref{2.1.1}). The tetrad and its dual give rise to the
completeness relations   
\begin{eqnarray}
g^{\mu\nu} &=& -u^{\mu} u^{\nu} 
+ \delta^{ab} \base{\mu}{a} \base{\nu}{b}, 
\nonumber \\ 
& & \label{2.1.3} \\ 
g_{\mu\nu} &=& -\base{0}{\mu} \base{0}{\nu} 
+ \delta_{ab}\, \base{a}{\mu} \base{b}{\nu} 
\nonumber
\end{eqnarray} 
for the metric and its inverse evaluated on the world line.  

The retarded coordinates are constructed with the help of a null
geodesic that links a given point $x$ to the world line. This geodesic
must be unique, and we thus restrict $x$ to be within the normal
convex neighborhood of $\gamma$. We denote by $\beta$ the unique,
future-directed null geodesic that goes from the world line to $x$,
and $x' \equiv z(u)$ is the point at which $\beta$ intersects
the world line; $u$ is the value of the proper-time parameter at this
point. To tensors at $x$ we shall assign the Greek indices $\alpha$,
$\beta$, $\gamma$, $\delta$, etc.; to tensors at $x'$ we shall assign
the indices $\alpha'$, $\beta'$, $\gamma'$, $\delta'$, and so on.     

From the tetrad $(u^{\alpha'},\base{\alpha'}{a})$ at $x'$ we construct
another tetrad $(\base{\alpha}{0},\base{\alpha}{a})$ at $x$ by
parallel transport on $\beta$. By raising the frame index and lowering
the vectorial index, we obtain also a dual tetrad at $x$:
$\base{0}{\alpha} = -g_{\alpha\beta} \base{\beta}{0}$ and 
$\base{a}{\alpha} = \delta^{ab} g_{\alpha\beta} \base{\beta}{b}$. The
metric at $x$ can then be expressed as 
\begin{equation}
g_{\alpha\beta} = -\base{0}{\alpha} \base{0}{\beta} + \delta_{ab} 
\base{a}{\alpha} \base{b}{\beta},   
\label{2.1.4}
\end{equation} 
and the parallel propagator \cite{synge:60} (also known as the
bivector of geodetic parallel displacement \cite{dewitt-brehme:60})
from $x'$ to $x$ is given by 
\begin{eqnarray} 
g^{\alpha}_{\ \alpha'}(x,x') &=& -\base{\alpha}{0} u_{\alpha'} 
+  \base{\alpha}{a} \base{a}{\alpha'},
\nonumber \\ 
& & \label{2.1.5} \\  
g^{\alpha'}_{\ \alpha}(x',x) &=& u^{\alpha'} \base{0}{\alpha} 
+ \base{\alpha'}{a} \base{a}{\alpha}. 
\nonumber 
\end{eqnarray}
This is defined such that if $A^\alpha$ is a vector that is parallel 
transported on $\beta$, then $A^\alpha(x) = g^\alpha_{\ \alpha'}(x,x') 
A^{\alpha'}(x')$ and $A^{\alpha'}(x') = g^{\alpha'}_{\ \alpha}(x',x) 
A^{\alpha}(x)$. Similarly, if $p_\alpha$ is a parallel-transported
dual vector, then $p_\alpha(x) = g^{\alpha'}_{\ \alpha}(x',x)
p_{\alpha'}(x')$ and $p_{\alpha'}(x') = g^{\alpha}_{\ \alpha'}(x,x') 
p_{\alpha}(x)$. 

The last ingredient we shall need is Synge's world function
$\sigma(z,x)$ \cite{synge:60} (also known as the biscalar of geodetic
interval \cite{dewitt-brehme:60}). This is defined as half the squared 
geodesic distance between the world-line point $z(\tau)$ and a
neighboring point $x$. The derivative of the world function with
respect to $z^\mu$ is denoted $\sigma_\mu(z,x)$; this is a vector at
$z$ (and a scalar at $x$) that is known to be tangent to the geodesic
linking $z$ and $x$. The derivative of $\sigma(z,x)$ with respect to
$x^\alpha$ is denoted $\sigma_\alpha(z,x)$; this vector at $x$ (and
scalar at $z$) is also tangent to the geodesic. We
use a similar notation for multiple derivatives; for example,
$\sigma_{\mu\alpha} \equiv \nabla_\alpha \nabla_\mu \sigma$ and
$\sigma_{\alpha\beta} \equiv \nabla_\beta \nabla_\alpha \sigma$, where 
$\nabla_\alpha$ denotes a covariant derivative at $x$ while
$\nabla_\mu$ indicates covariant differentiation at $z$.    

The vector $-\sigma^\mu(z,x)$ can be thought of as a separation vector
between $x$ and $z$, pointing from the world line to $x$. When $x$ is
close to $\gamma$, $-\sigma^\mu(z,x)$ is small and can be used to
express bitensors in terms of ordinary tensors at
$z$ \cite{synge:60, dewitt-brehme:60}. For example, 
\begin{eqnarray} 
\sigma_{\mu\nu} &=& g_{\mu\nu} - \frac{1}{3} R_{\mu\lambda\nu\rho}
\sigma^\lambda \sigma^\rho + \cdots, 
\label{2.1.6} \\ 
\sigma_{\mu\alpha} &=& -g^\nu_{\ \alpha} \Bigl(g_{\mu\nu} 
+ \frac{1}{6} R_{\mu\lambda\nu\rho} \sigma^\lambda \sigma^\rho  
+ \cdots \Bigr),
\label{2.1.7} \\ 
\sigma_{\alpha\beta} &=& g^\mu_{\ \alpha} g^\nu_{\ \beta} \Bigr( 
g_{\mu\nu} - \frac{1}{3} R_{\mu\lambda\nu\rho} \sigma^\lambda
\sigma^\rho + \cdots \Bigr), 
\label{2.1.8}
\end{eqnarray} 
where $g^\mu_{\ \alpha} \equiv g^\mu_{\ \alpha}(z,x)$ is the parallel
propagator and $R_{\mu\lambda\nu\rho}$ is the Riemann tensor evaluated
on the world line. 

\subsection{Definition of the retarded coordinates} 

The quasi-Cartesian version of the retarded coordinates are defined by    
\begin{equation}
\hat{x}^0 = u, \qquad
\hat{x}^a = -\base{a}{\alpha'}(x') \sigma^{\alpha'}(x,x'), \qquad  
\sigma(x,x') = 0.   
\label{2.2.1}
\end{equation}
The last statement indicates that $x' \equiv z(u)$ and $x$ are linked
by a null geodesic. \footnote{A similar definition can be given for
Fermi normal coordinates \cite{synge:60, manasse-misner:63}. Here 
$x$ is linked to $\gamma$ by a spacelike geodesic that
intersects the world line orthogonally. The intersection point is such
that $\sigma_\mu(x,z) u^\mu = 0$, and this replaces the last condition
of Eq.~(\ref{2.2.1}). The other relations are unchanged.}      

From the fact that $\sigma^{\alpha'}$ is a null vector we obtain    
\begin{equation} 
r \equiv (\delta_{ab} \hat{x}^a \hat{x}^b)^{1/2} 
= u_{\alpha'} \sigma^{\alpha'}, 
\label{2.2.2}
\end{equation} 
and $r$ is a positive quantity by virtue of the fact that $\beta$ is a  
future-directed null geodesic --- this makes $\sigma^{\alpha'}$
past-directed. In flat spacetime, $\sigma^{\alpha'} =
-(x-x')^{\alpha}$, and in a Lorentz frame that is momentarily comoving 
with the world line, $r = t-t' > 0$; with the speed of light set equal
to unity, $r$ is also the spatial distance between $x'$ and $x$ as
measured in this frame. This gives us an interpretation of $r = 
u_{\alpha'} \sigma^{\alpha'}$ as a {\it retarded distance} between $x$
and the world line, and we shall keep this interpretation even in
curved spacetime. The claim that $r$ gives a measure of distance
between $x'$ and $x$ will be substantiated in Sec.~II C.   

Another consequence of Eq.~(\ref{2.2.1}) is that 
\begin{equation}
\sigma^{\alpha'} = -r \bigl( u^{\alpha'} + \Omega^a \base{\alpha'}{a}
\bigr), 
\label{2.2.3}
\end{equation}
where $\Omega^a \equiv \hat{x}^a/r$ is a frame vector that satisfies  
$\delta_{ab} \Omega^a \Omega^b = 1$. 

A straightforward calculation reveals that under a displacement of the  
point $x$, the retarded coordinates change according to  
\begin{eqnarray} 
d u &=& -k_\alpha\, d x^\alpha, 
\label{2.2.4} \\ 
d \hat{x}^a &=& - \bigl( r a^a - \omega^a_{\ b} \hat{x}^b  
+ \base{a}{\alpha'} \sigma^{\alpha'}_{\ \beta'} u^{\beta'} 
\bigr)\, du 
\nonumber \\ & & \mbox{}
- \base{a}{\alpha'} \sigma^{\alpha'}_{\ \beta}\, d x^\beta,
 \label{2.2.5}
\end{eqnarray}
where $k_\alpha \equiv \sigma_{\alpha}/r$ is a future-directed null
vector at $x$ that is tangent to the geodesic $\beta$. To obtain these  
results we must keep in mind that a displacement of $x$ typically
induces a simultaneous displacement of $x'$, because the new points  
$x + \delta x$ and $x' + \delta x'$ must also be linked by a null 
geodesic. We therefore have $0 = \sigma(x+\delta x, x' + \delta x')  
= \sigma_{\alpha}\, \delta x^\alpha + \sigma_{\alpha'}\, 
\delta x^{\alpha'}$, and Eq.~(\ref{2.2.4}) follows from the fact that
a displacement along the world line is described by 
$\delta x^{\alpha'} = u^{\alpha'}\, \delta u$.  

\subsection{Retarded distance; null vector field}  

If we keep $x'$ linked to $x$ by the relation $\sigma(x,x') = 0$, then
\begin{equation}
r(x) = \sigma_{\alpha'}(x,x') u^{\alpha'}(x') 
\label{2.3.1}
\end{equation}
can be viewed as an ordinary scalar field defined in a neighborhood 
of $\gamma$. We can compute the gradient of $r$ by finding how $r$ 
changes under a displacement of $x$ (which again induces a
displacement of $x'$). The result is 
\begin{equation} 
\partial_\beta r = - \bigl( \sigma_{\alpha'} a^{\alpha'} +
\sigma_{\alpha'\beta'} u^{\alpha'} u^{\beta'} \bigr) k_\beta +
\sigma_{\alpha'\beta} u^{\alpha'}.
\label{2.3.2}
\end{equation} 

Similarly, we can view 
\begin{equation} 
k^{\alpha}(x) = \frac{\sigma^{\alpha}(x,x')}{r(x)} 
\label{2.3.3}
\end{equation}
as an ordinary vector field, which is tangent to the congruence of 
null geodesics that emanate from $x'$. It is easy to check that this 
vector satisfies the identities 
\begin{equation} 
\sigma_{\alpha\beta} k^\beta = k_\alpha, \qquad 
\sigma_{\alpha'\beta} k^\beta = \frac{\sigma_{\alpha'}}{r}, 
\label{2.3.4}
\end{equation}
from which we also obtain $\sigma_{\alpha'\beta} u^{\alpha'} k^\beta 
= 1$. From this last result and Eq.~(\ref{2.3.2}) we deduce the
important relation  
\begin{equation}
k^\alpha \partial_\alpha r = 1. 
\label{2.3.5}
\end{equation} 
In addition, combining the general statement $\sigma^{\alpha} =  
-g^{\alpha}_{\ \alpha'} \sigma^{\alpha'}$ with Eq.~(\ref{2.2.3}) gives 
\begin{equation} 
k^\alpha = g^{\alpha}_{\ \alpha'} \bigl( u^{\alpha'} + \Omega^a
\base{\alpha'}{a} \bigr); 
\label{2.3.6}
\end{equation} 
the vector at $x$ is therefore obtained by parallel transport of 
$u^{\alpha'} + \Omega^a \base{\alpha'}{a}$ on $\beta$. From this and
Eq.~(\ref{2.1.4}) we get the alternative expression 
\begin{equation} 
k^\alpha = \base{\alpha}{0} + \Omega^a \base{\alpha}{a},  
\label{2.3.7}
\end{equation} 
which confirms that $k^\alpha$ is a future-directed null vector field 
(recall that $\Omega^a = \hat{x}^a/r$ is a unit frame vector).    

The covariant derivative of $k_\alpha$ can be computed by finding  
how the vector changes under a displacement of $x$. (It is in fact
easier to first calculate how $r k_\alpha$ changes, and then
substitute our previous expression for $\partial_\beta r$.) The result
is   
\begin{eqnarray} 
r k_{\alpha;\beta} &=& \sigma_{\alpha\beta} 
- k_{\alpha} \sigma_{\beta \gamma'} u^{\gamma'} 
- k_{\beta} \sigma_{\alpha \gamma'} u^{\gamma'} 
\nonumber \\ & & \mbox{} 
+ \bigl( \sigma_{\alpha'} a^{\alpha'} + \sigma_{\alpha'\beta'} 
u^{\alpha'} u^{\beta'} \bigr) k_{\alpha} k_{\beta}. 
\label{2.3.8}
\end{eqnarray} 
From this we infer that $k^\alpha$ satisfies the geodesic equation in 
affine-parameter form, $k^\alpha_{\ ;\beta} k^\beta = 0$, and
Eq.~(\ref{2.3.5}) informs us that the affine parameter is in fact
$r$. A displacement along a member of the congruence is therefore 
described by $dx^\alpha = k^\alpha\, dr$. Specializing to retarded 
coordinates, and using Eqs.~(\ref{2.2.4}), (\ref{2.2.5}), and
(\ref{2.3.4}), we find that this statement becomes $du = 0$ and
$d\hat{x}^a = (\hat{x}^a/r)\, dr$, which integrate to $u =
\mbox{constant}$ and $\hat{x}^a = r \Omega^a$, respectively, with
$\Omega^a$ representing a constant unit vector. We have found that
the congruence of null geodesics emanating from $x'$ is described by    
\begin{equation}
u = \mbox{constant}, \qquad
\hat{x}^a = r \Omega^a(\theta^A)
\label{2.3.9}
\end{equation} 
in the retarded coordinates. Here, the two angles $\theta^A$ ($A = 1, 
2$) serve to parameterize the unit vector $\Omega^a$, which is
independent of $r$.       

Equation (\ref{2.3.8}) also implies that the expansion of the
congruence is given by 
\begin{equation} 
\Theta = k^\alpha_{\ ;\alpha} = \frac{\sigma^\alpha_{\ \alpha} 
- 2}{r}.
\label{2.3.10}
\end{equation} 
Using the expansion for $\sigma^\alpha_{\ \alpha}$ given by
Eq.~(\ref{2.1.8}), we find that this becomes $r\Theta = 2  
- \frac{1}{3} R_{\alpha'\beta'} \sigma^{\alpha'} \sigma^{\beta'}  
+ O(r^3)$, or 
\begin{equation}
r \Theta =  
2 - \frac{1}{3} r^2 \bigl( R_{00} + 2 R_{0a} \Omega^a + R_{ab}
\Omega^a \Omega^b \bigr) + O(r^3)
\label{2.3.11} 
\end{equation}
after using Eq.~(\ref{2.2.3}). Here, 
$R_{00} = R_{\alpha'\beta'} u^{\alpha'} u^{\beta'}$,
$R_{0a} = R_{\alpha'\beta'} u^{\alpha'} \base{\beta'}{a}$, and  
$R_{ab} = R_{\alpha'\beta'} \base{\alpha'}{a} \base{\beta'}{b}$ 
are the frame components of the Ricci tensor evaluated at $x'$. This 
result confirms that the congruence is singular at $r=0$, because
$\Theta$ diverges as $2/r$ in this limit; the caustic coincides with
the point $x'$.      

Finally, we infer from Eq.~(\ref{2.3.8}) that $k^{\alpha}$ is 
hypersurface orthogonal. This, together with the property that
$k^\alpha$ satisfies the geodesic equation in affine-parameter form, 
implies that there exists a scalar field $u(x)$ such that  
\begin{equation}
k_\alpha = -\partial_\alpha u. 
\label{2.3.12}
\end{equation} 
This scalar field was already identified in Eq.~(\ref{2.2.4}): it is  
numerically equal to the proper-time parameter of the world line at
$x'$. We may thus conclude that the geodesics to which $k^\alpha$ is
tangent are the generators of the null cone $u = \mbox{constant}$. As 
Eq.~(\ref{2.3.9}) indicates, a specific generator is selected by 
choosing a direction $\Omega^a$ (which can be parameterized by two
angles $\theta^A$), and $r$ is an affine parameter on each
generator. The geometrical meaning of the retarded coordinates is now
completely clear, and $r$ is recognized as a meaningful measure of
distance between $x$ and the world line. The construction is
illustrated in Fig.~1. 

\begin{figure}
\includegraphics[angle=0,scale=0.4]{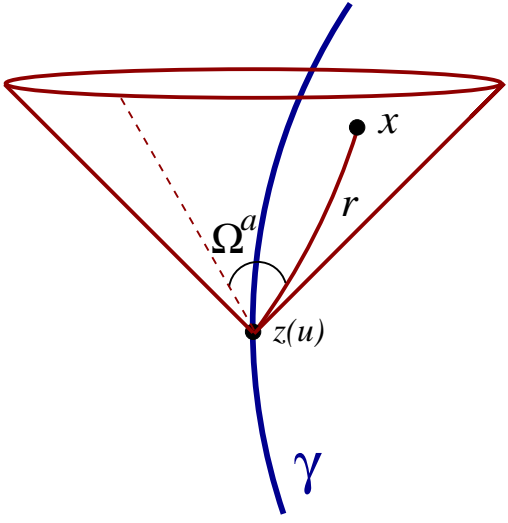} 
\caption{Retarded coordinates of a point $x$ relative to a world 
line $\gamma$. The retarded time $u$ selects a particular null cone, 
the unit vector $\Omega^a \equiv \hat{x}^a/r$ selects a particular
generator of this null cone, and the retarded distance $r$ selects a 
particular point on this generator.} 
\end{figure} 

\subsection{Frame components of tensors on the world line}   

The metric at $x$ in the retarded coordinates will be expressed in 
terms of frame components of vectors and tensors evaluated on
$\gamma$. For example, if $a^{\alpha'}$ is the acceleration vector at
$x'$, then as we have seen,  
\begin{equation}
a_a(u) = a_{\alpha'}\, \base{\alpha'}{a}
\label{2.4.1}
\end{equation}
are the frame components of the acceleration at proper time $u$.   

Similarly,   
\begin{eqnarray}
R_{a0b0}(u) &=& R_{\alpha'\gamma'\beta'\delta'}\, \base{\alpha'}{a}  
u^{\gamma'} \base{\beta'}{b} u^{\delta'}, 
\nonumber \\ 
R_{a0bd}(u) &=& R_{\alpha'\gamma'\beta'\delta'}\, \base{\alpha'}{a} 
u^{\gamma'} \base{\beta'}{b} \base{\delta'}{d}, 
\label{2.4.2} \\ 
R_{acbd}(u) &=& R_{\alpha'\gamma'\beta'\delta'}\, \base{\alpha'}{a} 
\base{\gamma'}{c} \base{\beta'}{b} \base{\delta'}{d} 
\nonumber 
\end{eqnarray}
are the frame components of the Riemann tensor evaluated on 
$\gamma$. From these we form the useful combinations  
\begin{eqnarray}
S_{ab} &=& R_{a0b0} + R_{a0bc} \Omega^c 
+ R_{b0ac} \Omega^c + R_{acbd} \Omega^c \Omega^d 
\nonumber \\
&=& S_{ba}, 
\label{2.4.3} \\
S_{a} &=& S_{ab}\Omega^b = R_{a0b0} \Omega^b - R_{ab0c} 
\Omega^b \Omega^c, 
\label{2.4.4} \\
S &=& S_{a} \Omega^a = R_{a0b0} \Omega^a \Omega^b, 
\label{2.4.5}
\end{eqnarray}
in which the quantities $\Omega^a \equiv \hat{x}^a / r$ depend on the   
angles $\theta^A$ only --- they are independent of $u$ and $r$.  

We have previously introduced the frame components of the Ricci tensor 
in Eq.~(\ref{2.3.11}). The identity  
\begin{equation}    
R_{00} + 2 R_{0a} \Omega^a + R_{ab} \Omega^a \Omega^b 
= \delta^{ab} S_{ab} - S
\label{2.4.6}
\end{equation}
follows easily from Eqs.~(\ref{2.4.3})--(\ref{2.4.5}) and the 
definition of the Ricci tensor. 

\subsection{Coordinate displacements near $\gamma$} 

The changes in the quasi-Cartesian retarded coordinates under a
displacement of $x$ are given by Eqs.~(\ref{2.2.4}) and
(\ref{2.2.5}). In these we substitute the expansions for
$\sigma_{\alpha'\beta'}$ and $\sigma_{\alpha'\beta}$ that appear in
Eqs.~(\ref{2.1.6}) and (\ref{2.1.7}), as well as Eqs.~(\ref{2.2.3})
and (\ref{2.3.6}). After a straightforward calculation, we obtain the
following expressions for the coordinate displacements:   
\begin{eqnarray}
du &=& \bigl( \base{0}{\alpha}\, dx^\alpha \bigr) 
- \Omega_a \bigl( \base{b}{\alpha}\, dx^\alpha \bigr),  
\label{2.5.1} \\ 
d\hat{x}^a &=& - \Bigl[ r a^a - r \omega^a_{\ b} \Omega^b 
+ \frac{1}{2} r^2 S^a + O(r^3) \Bigr]
\bigl( \base{0}{\alpha}\, dx^\alpha \bigr) 
\nonumber \\ & & \mbox{} 
+ \Bigl[ \delta^a_{\ b} + \Bigl( r a^a - r \omega^a_{\ c}
\Omega^c + \frac{1}{3} r^2 S^a \Bigr) \Omega_b 
\nonumber \\ & & \mbox{} 
+ \frac{1}{6} r^2 S^a_{\ b} + O(r^3) \Bigr] 
\bigl( \base{b}{\alpha}\, dx^\alpha \bigr).    
\label{2.5.2}
\end{eqnarray} 
Notice that the result for $du$ is exact, but that $d\hat{x}^a$ is
expressed as an expansion in powers of $r$.  

These results can also be expressed in the form of gradients of the
retarded coordinates: 
\begin{eqnarray} 
\partial_\alpha u &=& \base{0}{\alpha} - \Omega_a \base{a}{\alpha},   
\label{2.5.3} \\ 
\partial_\alpha \hat{x}^a &=& - \Bigl[ r a^a - r \omega^a_{\ b}
\Omega^b + \frac{1}{2} r^2 S^a + O(r^3) \Bigr] \base{0}{\alpha} 
\nonumber \\ & & \mbox{} 
+ \Bigl[ \delta^a_{\ b} + \Bigl( r a^a - r \omega^a_{\ c}
\Omega^c + \frac{1}{3} r^2 S^a \Bigr) \Omega_b 
\nonumber \\ & & \mbox{} 
+ \frac{1}{6} r^2 S^a_{\ b} + O(r^3) \Bigr] \base{b}{\alpha}. 
\label{2.5.4}
\end{eqnarray} 
Notice that Eq.~(\ref{2.5.3}) follows immediately from
Eqs.~(\ref{2.3.7}) and (\ref{2.3.12}). From Eq.~(\ref{2.5.4}) and the  
identity $\partial_\alpha r = \Omega_a \partial_\alpha \hat{x}^a$ we
also infer 
\begin{eqnarray} 
\partial_\alpha r &=& - \Bigl[ r a_a \Omega^a + \frac{1}{2} r^2 S  
+ O(r^3) \Bigr] \base{0}{\alpha} 
\nonumber \\ & & \mbox{} 
+ \Bigl[ \Bigl(1 + r a_b \Omega^b + \frac{1}{3} r^2 S \Bigr) \Omega_a  
\nonumber \\ & & \mbox{} 
+ \frac{1}{6} r^2 S_a + O(r^3) \Bigr] \base{a}{\alpha}, 
\label{2.5.5}
\end{eqnarray} 
where we have used the facts that $S_a = S_{ab} \Omega^b$ and $S = S_a 
\Omega^a$; see Eqs.~(\ref{2.4.4}) and (\ref{2.4.5}). It may be checked
that Eq.~(\ref{2.5.5}) agrees with Eq.~(\ref{2.3.2}).  

\subsection{Metric near $\gamma$}   

It is straightforward to invert the relations of 
Eqs.~(\ref{2.5.1}) and (\ref{2.5.2}) and solve for $\base{0}{\alpha}\,  
dx^\alpha$ and $\base{a}{\alpha}\, dx^\alpha$. The results are  
\begin{eqnarray} 
\base{0}{\alpha}\, dx^\alpha &=& \Bigl[ 1 + r a_a \Omega^a 
+ \frac{1}{2} r^2 S + O(r^3) \Bigr]\, du 
\nonumber \\ & & \mbox{} 
+ \Bigl[ \Bigl( 1 + \frac{1}{6} r^2 S \Bigr) \Omega_a 
\nonumber \\ & & \mbox{} 
- \frac{1}{6} r^2 S_a + O(r^3) \Bigr]\, d\hat{x}^a, 
\label{2.6.1} \\ 
\base{a}{\alpha}\, dx^\alpha &=& \Bigl[ r \bigl(a^a - \omega^a_{\ b} 
\Omega^b \bigr) + \frac{1}{2} r^2 S^a + O(r^3) \Bigr]\, du 
\nonumber \\ & & \mbox{} 
+ \Bigl[ \delta^a_{\ b} - \frac{1}{6} r^2 S^a_{\ b} 
\nonumber \\ & & \mbox{} 
+ \frac{1}{6} r^2 S^a \Omega_b + O(r^3) \Bigr]\, d\hat{x}^b.
\label{2.6.2}
\end{eqnarray}  
These relations, when specialized to the retarded coordinates, give us  
the components of the dual tetrad $(\base{0}{\alpha},
\base{a}{\alpha})$ at $x$. The metric is then computed by using the  
completeness relations of Eq.~(\ref{2.1.3}). We find 
\begin{eqnarray}
g_{uu} &=& - \bigl( 1 + r a_a \Omega^a \bigr)^2 + r^2 
\bigl(a_a - \omega_{ab} \Omega^b \bigr) \bigl(a^a - \omega^a_{\ c} 
\Omega^c \bigr) 
\nonumber \\ & & \mbox{} 
- r^2 S + O(r^3),  
\label{2.6.3} \\ 
g_{ua} &=& -\Bigl( 1 + r a_b \Omega^b + \frac{2}{3} r^2 S \Bigr) 
\Omega_a + r \bigl(a_a - \omega_{ab} \Omega^b \bigr) 
\nonumber \\ & & \mbox{} 
+ \frac{2}{3} r^2 S_a + O(r^3), 
\label{2.6.4} \\
g_{ab} &=& \delta_{ab} - \Bigl( 1 + \frac{1}{3} r^2 S \Bigr) \Omega_a 
\Omega_b - \frac{1}{3} r^2 S_{ab} 
\nonumber \\ & & \mbox{} 
+ \frac{1}{3} r^2 \bigl( S_a \Omega_b + \Omega_a S_b \bigr) + O(r^3).   
\label{2.6.5}
\end{eqnarray} 
We see that the metric possesses a directional ambiguity on the world
line: the metric at $r=0$ still depends on the vector $\Omega^a =
\hat{x}^a/r$ that specifies the direction to the point $x$. The
retarded coordinates are therefore singular on the world line, and
tensor components cannot be defined on $\gamma$. Because we are
working with {\it frame components} of tensors, this poses no
particular difficulty. 

By setting $S_{ab} = S_a = S = 0$ in Eqs.~(\ref{2.6.3})--(\ref{2.6.5})   
we obtain the metric of flat spacetime in the retarded
coordinates. This we express as  
\begin{eqnarray} 
\eta_{uu} &=& - \bigl( 1 + r a_a \Omega^a \bigr)^2 
+ r^2 \bigl(a_a - \omega_{ab} \Omega^b \bigr) 
\bigl(a^a - \omega^a_{\ c} \Omega^c \bigr), \nonumber \\   
\eta_{ua} &=& - \bigl( 1 + r a_b \Omega^b \bigr) \Omega_a 
+ r \bigl(a_a - \omega_{ab} \Omega^b \bigr),
\label{2.6.6} \\ 
\eta_{ab} &=& \delta_{ab} - \Omega_a \Omega_b, \nonumber 
\end{eqnarray}
and we see that the directional ambiguity persists. This should not
come as a surprise: the ambiguity is present even when $a_a =
\omega_{ab} = 0$, and is generated simply by performing the coordinate
transformation $u = t - \sqrt{x^2+y^2+z^2}$. The retarded coordinates
are therefore necessarily singular at the world line. But in spite of
the directional ambiguity, the metric of flat spacetime has a unit 
determinant everywhere, and it is easily inverted:    
\begin{eqnarray}
\eta^{uu} &=& 0, 
\nonumber \\ 
\eta^{ua} &=& -\Omega^a, 
\label{2.6.7} \\ 
\eta^{ab} &=& \delta^{ab} + r \bigl(a^a - \omega^a_{\ c} 
\Omega^c \bigr) \Omega^b + r \Omega^a \bigl(a^b 
- \omega^b_{\ c} \Omega^c \bigr).  
\nonumber
\end{eqnarray}
The inverse metric also is ambiguous on the world line.   

To invert the curved-spacetime metric of
Eqs.~(\ref{2.6.3})--(\ref{2.6.5}) we express it as $g_{\alpha\beta} = 
\eta_{\alpha\beta} + h_{\alpha\beta} + O(r^3)$ and treat
$h_{\alpha\beta} = O(r^2)$ as a perturbation. The inverse metric is
then $g^{\alpha\beta} = \eta^{\alpha\beta} - \eta^{\alpha\gamma}
\eta^{\beta\delta} h_{\gamma \delta} + O(r^3)$, or  
\begin{eqnarray} 
g^{uu} &=& 0, 
\label{2.6.8} \\
g^{ua} &=& -\Omega^a, 
\label{2.6.9} \\
g^{ab} &=& \delta^{ab} + r \bigl(a^a - \omega^a_{\ c}
\Omega^c \bigr) \Omega^b + r \Omega^a \bigl(a^b - \omega^b_{\ c}
\Omega^c \bigr) 
\nonumber \\ & & \hspace*{-10pt} \mbox{} 
+ \frac{1}{3} r^2 S^{ab} + \frac{1}{3} r^2 
\bigl( S^a \Omega^b + \Omega^a S^b \bigr) + O(r^3).  
\label{2.6.10}
\end{eqnarray} 
The results for $g^{uu}$ and $g^{ua}$ are exact, and they follow from   
the general relations $g^{\alpha\beta} (\partial_\alpha u)
(\partial_\beta u) = 0$ and $g^{\alpha\beta} (\partial_\alpha u)
(\partial_\beta r) = -1$ that are derived from Eqs.~(\ref{2.3.5}) and 
(\ref{2.3.12}).  

The metric determinant is computed from $\sqrt{-g} = 1 + \frac{1}{2}  
\eta^{\alpha\beta} h_{\alpha\beta} + O(r^3)$, which gives 
\begin{eqnarray}
\sqrt{-g} &=& 1 - \frac{1}{6} r^2 \bigl( \delta^{ab} S_{ab} 
- S \bigr) + O(r^3) 
\nonumber \\ 
&=& 1 - \frac{1}{6} r^2 \bigl( R_{00} + 2 R_{0a} \Omega^a 
+ R_{ab} \Omega^a \Omega^b \bigr) 
\nonumber \\ & & \mbox{} 
+ O(r^3), 
\label{2.6.11}
\end{eqnarray}
where we have substituted the identity of
Eq.~(\ref{2.4.6}). Comparison with Eq.~(\ref{2.3.11}) gives us
the interesting relation $\sqrt{-g} = \frac{1}{2} r \Theta + O(r^3)$, 
where $\Theta$ is the expansion of the generators of the null cones 
$u = \mbox{constant}$.  

\subsection{Transformation to angular coordinates} 

Because the frame vector $\Omega^a = \hat{x}^a/r$ satisfies
$\delta_{ab} \Omega^a \Omega^b = 1$, it can be parameterized by two
angles $\theta^A$. A canonical choice for the parameterization is
$\Omega^a = (\sin\theta \cos\phi, \sin\theta \sin\phi,
\cos\theta)$. It is then convenient to perform a coordinate
transformation from $\hat{x}^a$ to $(r,\theta^A)$, using the relations
$\hat{x}^a = r \Omega^a(\theta^A)$. (Recall from Sec.~II C that the
angles $\theta^A$ are constant on the generators of the null cones $u
= \mbox{constant}$, and that $r$ is an affine parameter on these
generators. The relations $\hat{x}^a = r \Omega^a$ therefore
describe the behavior of the generators.) The differential form of 
the coordinate transformation is     
\begin{equation}
d\hat{x}^a = \Omega^a\, dr + r \Omega^a_A\, d\theta^A, 
\label{2.7.1}
\end{equation}
where the transformation matrix 
\begin{equation} 
\Omega^a_A \equiv \frac{\partial \Omega^a}{\partial \theta^A}
\label{2.7.2}
\end{equation} 
satisfies the identity $\Omega_a \Omega^a_A = 0$. 

We introduce the quantities 
\begin{equation}
\Omega_{AB} = \delta_{ab} \Omega^a_A \Omega^b_B, 
\label{2.7.3}
\end{equation}
which act as a (nonphysical) metric in the subspace spanned by the 
angular coordinates. In the canonical parameterization, $\Omega_{AB} =  
\mbox{diag}(1,\sin^2\theta)$. We use the inverse of $\Omega_{AB}$, 
denoted $\Omega^{AB}$, to raise upper-case Latin indices. We then
define the new object 
\begin{equation}
\Omega^A_a = \delta_{ab} \Omega^{AB} \Omega^b_B 
\label{2.7.4}
\end{equation} 
which satisfies the identities 
\begin{equation}
\Omega^A_a \Omega^a_B = \delta^A_B, \qquad
\Omega^a_A \Omega^A_b = \delta^a_{\ b} - \Omega^a \Omega_b. 
\label{2.7.5}
\end{equation} 
The second result follows from the fact that both sides are
symmetric in $a$ and $b$, orthogonal to $\Omega_a$ and  
$\Omega^b$, and have the same trace.  

From the preceding results we establish that the transformation from 
$\hat{x}^a$ to $(r,\theta^A)$ is accomplished by 
\begin{equation}
\frac{\partial \hat{x}^a}{\partial r} = \Omega^a, \qquad
\frac{\partial \hat{x}^a}{\partial \theta^A} = r \Omega^a_A, 
\label{2.7.6}
\end{equation}
while the transformation from $(r,\theta^A)$ to $\hat{x}^a$ is
accomplished by 
\begin{equation}
\frac{\partial r}{\partial \hat{x}^a} = \Omega_a, \qquad 
\frac{\partial \theta^A}{\partial \hat{x}^a} = \frac{1}{r}\Omega^A_a.  
\label{2.7.7}
\end{equation}
With these rules it is easy to show that in the 
angular coordinates, the metric takes the form of 
\begin{equation}
ds^2 = g_{uu}\, du^2 - 2\, dudr + 2 g_{uA}\, du d\theta^A 
+ g_{AB}\, d\theta^A d\theta^B,
\label{2.7.8}
\end{equation} 
with  
\begin{eqnarray}
g_{uu} &=& - \bigl( 1 + r a_a \Omega^a \bigr)^2 
+ r^2 \bigl(a_a - \omega_{ab} \Omega^b \bigr) 
\bigl(a^a - \omega^a_{\ c} \Omega^c \bigr) 
\nonumber \\ & & \mbox{} 
- r^2 S + O(r^3),  
\label{2.7.9} \\
g_{uA} &=& r \Bigl[ r \bigl(a_a - \omega_{ab} \Omega^b \bigr) 
+ \frac{2}{3} r^2 S_a + O(r^3) \Bigr] \Omega^a_A, 
\label{2.7.10} \\ 
g_{AB} &=& r^2 \Bigl[ \Omega_{AB} - \frac{1}{3} r^2 S_{ab} \Omega^a_A
\Omega^b_B + O(r^3) \Bigr]. 
\label{2.7.11}
\end{eqnarray} 
The results $g_{ru} = -1$, $g_{rr} = 0$, and $g_{rA} = 0$ are exact, 
and they follow from the fact that in the retarded coordinates,
$k_\alpha\, dx^\alpha = - du$ and $k^\alpha \partial_\alpha =
\partial_r$.   

The nonvanishing components of the inverse metric are  
\begin{eqnarray}
g^{ur} &=& -1, 
\label{2.7.12} \\ 
g^{rr} &=& 1 + 2r a_a \Omega^a + r^2 S + O(r^3),   
\label{2.7.13} \\ 
g^{rA} &=& \frac{1}{r} \Bigl[ r \bigl(a^a - \omega^a_{\ b} \Omega^b
\bigr) + \frac{2}{3}r^2 S^a 
\nonumber \\ & & \mbox{} 
+ O(r^3) \Bigr] \Omega^A_a, 
\label{2.7.14} \\
g^{AB} &=& \frac{1}{r^2} \Bigl[ \Omega^{AB} + \frac{1}{3} r^2 S^{ab}
\Omega^A_a \Omega^B_b + O(r^3) \Bigr].  
\label{2.7.15}
\end{eqnarray} 
The results $g^{uu} = 0$, $g^{ur} = -1$, and $g^{uA} = 0$ are exact, 
and they follow from the same reasoning as before.  

Finally, we note that in the angular coordinates, the metric
determinant is given by  
\begin{equation}
\sqrt{-g} = r^2 \sqrt{\Omega} \Bigl[ 1 - \frac{1}{6} r^2 \bigl( R_{00}
+ 2 R_{0a} \Omega^a + R_{ab} \Omega^a \Omega^b \bigr) + O(r^3) \Bigr], 
\label{2.7.16}
\end{equation}
where $\Omega$ is the determinant of $\Omega_{AB}$; in the canonical 
parameterization, $\sqrt{\Omega} = \sin\theta$. 
 
\subsection{Specialization to $a^\mu = 0 = R_{\mu\nu}$}  

In this subsection we specialize our previous results to a situation  
where $\gamma$ is a geodesic on which the Ricci tensor vanishes. We
therefore set $a^\mu = 0 = R_{\mu\nu}$ everywhere on $\gamma$, and for
simplicity we also set $\omega_{ab}$ to zero.         

It is known that when the Ricci tensor vanishes, the Riemann tensor
can be decomposed in terms of a timelike vector $u^\alpha$ and two
symmetric-tracefree, spatial tensors ${\cal E}_{\alpha\beta}$ and 
${\cal B}_{\alpha\beta}$ (see, for example,
Ref.~\cite{thorne-hartle:85}). In terms of frame components we have    
\begin{eqnarray} 
R_{a0b0}(u) &=& {\cal E}_{ab} 
\nonumber \\ 
R_{a0bc}(u) &=& \varepsilon_{bcd} {\cal B}^d_{\ a},
\label{2.8.1} \\ 
R_{acbd}(u) &=& \delta_{ab} {\cal E}_{cd} 
+ \delta_{cd} {\cal E}_{ab} - \delta_{ad} {\cal E}_{bc} 
- \delta_{bc} {\cal E}_{ad}, 
\nonumber 
\end{eqnarray}
where ${\cal E}_{ab}$ and ${\cal B}_{ab}$ depend on $u$, are such that
${\cal E}_{ba} = {\cal E}_{ab}$, $\delta^{ab} {\cal E}_{ab} = 0$, 
${\cal B}_{ba} = {\cal B}_{ab}$, $\delta^{ab} {\cal B}_{ab} = 0$, 
and $\varepsilon_{abc}$ is the three-dimensional permutation
symbol. These relations can be substituted into
Eqs.~(\ref{2.4.3})--(\ref{2.4.5}) to give  
\begin{eqnarray}
S_{ab} &=& 2 {\cal E}_{ab} 
- \Omega_a {\cal E}_{bc} \Omega^c 
- \Omega_b {\cal E}_{ac} \Omega^c 
+ \delta_{ab} {\cal E}_{bc} \Omega^c \Omega^d 
\nonumber \\ & & \mbox{} 
+ \varepsilon_{acd} \Omega^c {\cal B}^d_{\ b} 
+ \varepsilon_{bcd} \Omega^c {\cal B}^d_{\ a}, 
\label{2.8.2} \\ 
S_{a} &=& {\cal E}_{ab} \Omega^b 
+ \varepsilon_{abc} \Omega^b {\cal B}^c_{\ d} \Omega^d, 
\label{2.8.3} \\ 
S &=& {\cal E}_{ab} \Omega^a \Omega^b.  
\label{2.8.4}
\end{eqnarray} 
In these expressions the dependence on retarded time $u$ is contained 
in ${\cal E}_{ab}$ and ${\cal B}_{ab}$, while the angular dependence
is encoded in the unit vector $\Omega^a$. 

It is convenient to introduce the irreducible quantities  
\begin{eqnarray}  
{\cal E}^* &=& {\cal E}_{ab} \Omega^a \Omega^b, 
\label{2.8.5} \\ 
{\cal E}^*_a &=& \bigl(\delta_a^{\ b} - \Omega_a \Omega^b \bigr) 
{\cal E}_{bc} \Omega^c,
\label{2.8.6} \\ 
{\cal E}^*_{ab} &=& 2 {\cal E}_{ab} 
- 2\Omega_a {\cal E}_{bc} \Omega^c  
- 2\Omega_b {\cal E}_{ac} \Omega^c
\nonumber \\ & & \mbox{} 
+ (\delta_{ab} + \Omega_a \Omega_b) {\cal E}^*, 
\label{2.8.7} \\ 
{\cal B}^*_a &=& \varepsilon_{abc} \Omega^b {\cal B}^c_{\ d} \Omega^d, 
\label{2.8.8} \\ 
{\cal B}^*_{ab} &=& 2\bigl(\delta_{(a}^{\ e} 
- \Omega_{(a}\Omega^e \bigr) \varepsilon_{b)cd} 
\Omega^c {\cal B}^d_{\ e}.  
\label{2.8.9}
\end{eqnarray}    
These are all orthogonal to $\Omega^a$: ${\cal E}^*_a \Omega^a =  
{\cal B}^*_a \Omega^a = 0$ and ${\cal E}^*_{ab} \Omega^b = 
{\cal B}^*_{ab} \Omega^b = 0$. In terms of these
Eqs.~(\ref{2.8.2})--(\ref{2.8.4}) become 
\begin{eqnarray} 
S_{ab} &=& {\cal E}_{ab}^* + \Omega_a {\cal E}^*_b + {\cal E}^*_a 
\Omega_b + \Omega_a \Omega_b {\cal E}^* 
\nonumber \\ & & \mbox{} 
+ {\cal B}^*_{ab} + \Omega_a {\cal B}^*_b + {\cal B}^*_a \Omega_b,  
\label{2.8.10} \\ 
S_{a} &=& {\cal E}^*_a + \Omega_a {\cal E}^* + {\cal B}^*_a, 
\label{2.8.11} \\ 
S &=& {\cal E}^*. 
\label{2.8.12}
\end{eqnarray} 

When Eqs.~(\ref{2.8.10})--(\ref{2.8.12}) are substituted into the
metric tensor of Eqs.~(\ref{2.6.3})--(\ref{2.6.5}) --- in which $a_a$ 
and $\omega_{ab}$ are both set equal to zero --- we obtain the compact 
expressions   
\begin{eqnarray}  
g_{uu} &=& - 1 - r^2 {\cal E}^* + O(r^3),
\label{2.8.13} \\ 
g_{ua} &=& -\Omega_a + \frac{2}{3} r^2 \bigl( {\cal E}^*_a 
+ {\cal B}^*_a \bigr) + O(r^3), 
\label{2.8.14} \\ 
g_{ab} &=& \delta_{ab} - \Omega_a \Omega_b - \frac{1}{3} r^2  
\bigl( {\cal E}^*_{ab} + {\cal B}^*_{ab} \bigr) 
\nonumber \\ & & \mbox{} 
+ O(r^3). 
\label{2.8.15}
\end{eqnarray}   
The metric becomes 
\begin{eqnarray} 
g_{uu} &=& - 1 - r^2 {\cal E}^* + O(r^3),
\label{2.8.16} \\ 
g_{ur} &=& - 1, 
\label{2.8.17} \\
g_{uA} &=& \frac{2}{3} r^3 \bigl( {\cal E}^*_A 
+ {\cal B}^*_A \bigr) + O(r^4), 
\label{2.8.18} \\ 
g_{AB} &=& r^2 \Omega_{AB} - \frac{1}{3} r^4  
\bigl( {\cal E}^*_{AB} + {\cal B}^*_{AB} \bigr) + O(r^5)  
\label{2.8.19}
\end{eqnarray}   
after transforming to angular coordinates using the rules of  
Eq.~(\ref{2.7.6}). Here we have introduced the projections 
\begin{eqnarray} 
{\cal E}^*_A &\equiv& {\cal E}^*_{a} \Omega^a_A 
= {\cal E}_{ab} \Omega^a_A \Omega^b, 
\label{2.8.20} \\
{\cal E}^*_{AB} &\equiv& {\cal E}^*_{ab} \Omega^a_A \Omega^b_B 
= 2 {\cal E}_{ab} \Omega^a_A \Omega^b_B + {\cal E}^* \Omega_{AB}, 
\label{2.8.21} \\
{\cal B}^*_A &\equiv& {\cal B}^*_{a} \Omega^a_A 
= \varepsilon_{abc} \Omega^a_A \Omega^b {\cal B}^c_{\ d} \Omega^d, 
\label{2.8.22} \\
{\cal B}^*_{AB} &\equiv& {\cal B}^*_{ab} \Omega^a_A \Omega^b_B 
= 2 \varepsilon_{acd} \Omega^c {\cal B}^d_{\ b} \Omega^{a}_{(A}  
\Omega^{b}_{B)}.  
\label{2.8.23}
\end{eqnarray} 
It may be noted that the inverse relations are ${\cal E}^*_a    
= {\cal E}^*_{A} \Omega^A_a$, ${\cal B}^*_a 
= {\cal B}^*_{A} \Omega^A_a$, ${\cal E}^*_{ab} 
= {\cal E}^*_{AB} \Omega^A_a \Omega^B_b$, and ${\cal B}^*_{ab}  
= {\cal B}^*_{AB} \Omega^A_a \Omega^B_b$, where $\Omega^A_a$ was
introduced in Eq.~(\ref{2.7.4}). 

The angular dependence of the quantities listed in
Eqs.~(\ref{2.8.20})--(\ref{2.8.23}) can be made more explicit by
expressing them in terms of scalar, vectorial, and tensorial spherical
harmonics. Let  
\begin{equation} 
Y^{\sf m} = \bigl\{ Y^{0}, Y^{1c}, Y^{1s}, Y^{2c}, Y^{2s}
\bigr\} 
\label{2.8.24}
\end{equation} 
be a set of real, unnormalized, spherical-harmonic functions of degree  
$l=2$; explicit expressions are provided in Appendix A. The numerical
part of the label ${\sf m}$ refers to the azimuthal index $m$ and the
letter indicates whether the function is proportional to $\cos(m\phi)$
or $\sin(m\phi)$. Vectorial harmonics are defined by   
\begin{equation} 
Y^{\sf m}_A = Y^{\sf m}_{:A}, \qquad 
X^{\sf m}_A = -\varepsilon_A^{\ B} Y^{\sf m}_{:B}, 
\label{2.8.25}
\end{equation} 
where a colon indicates covariant differentiation with respect to a
connection compatible with $\Omega_{AB}$, and $\varepsilon_{AB}$ is
the two-dimensional Levi-Civita tensor. The vectorial harmonics
$Y^{\rm m}_A$ have even parity, while $X^{\rm m}_A$ have odd
parity. Tensorial harmonics are defined by  
\begin{equation} 
Y^{\sf m} \Omega_{AB}, \quad 
Y^{\sf m}_{AB} = Y^{\sf m}_{:AB}, \quad 
X^{\sf m}_{AB} = -X^{\sf m}_{(A:B)}; 
\label{2.8.26}
\end{equation} 
the harmonics $Y^{\sf m} \Omega_{AB}$ and $Y^{\sf m}_{AB}$ have even  
parity, while $X^{\sf m}_{AB}$ have odd parity. Apart from notation
and normalization, these definitions agree with those of Regge and
Wheeler \cite{regge-wheeler:57}, and explicit expressions appear in
Appendix A.   

We define the harmonic components ${\cal E}_{\sf m}$ of the tensor
${\cal E}_{ab}$ with the relations  
\begin{eqnarray} 
{\cal E}_0 &=& {\cal E}_{33} = 
-\bigl({\cal E}_{11} + {\cal E}_{22}\bigr), 
\nonumber \\ 
{\cal E}_{1c} &=& 2 {\cal E}_{13}, 
\nonumber \\ 
{\cal E}_{1s} &=& 2 {\cal E}_{23}, 
\label{2.8.27} \\ 
{\cal E}_{2c} &=& \frac{1}{2} \bigl({\cal E}_{11} - 
{\cal E}_{22}\bigr), 
\nonumber \\ 
{\cal E}_{2s} &=& 2 {\cal E}_{12}. 
\nonumber 
\end{eqnarray}
Similarly, we define the harmonic components ${\cal B}_{\sf m}$ of the
tensor ${\cal B}_{ab}$ by 
\begin{eqnarray} 
{\cal B}_0 &=& {\cal B}_{33} =
-\bigl({\cal B}_{11} + {\cal B}_{22}\bigr), 
\nonumber \\ 
{\cal B}_{1c} &=& 2 {\cal B}_{13}, 
\nonumber \\ 
{\cal B}_{1s} &=& 2 {\cal B}_{23}, 
\label{2.8.28} \\ 
{\cal B}_{2c} &=& \frac{1}{2} \bigl({\cal B}_{11} - 
{\cal B}_{22}\bigr), 
\nonumber \\ 
{\cal B}_{2s} &=& 2 {\cal B}_{12}. 
\nonumber 
\end{eqnarray}
It is then straightforward to prove that 
Eqs.~(\ref{2.8.20})--(\ref{2.8.23}) are equivalent to 
\begin{eqnarray} 
{\cal E}^* &=& \sum_{\sf m} {\cal E}_{\sf m} Y^{\sf m}, 
\nonumber \\ 
{\cal E}^*_A &=& \frac{1}{2} \sum_{\sf m} {\cal E}_{\sf m} 
Y^{\sf m}_A,
\nonumber \\ 
{\cal E}^*_{AB} &=& \sum_{\sf m} {\cal E}_{\sf m} 
\bigl( Y^{\sf m}_{AB} + 3 Y^{\sf m} \Omega_{AB} \bigr), 
\label{2.8.29} \\ 
{\cal B}^*_A &=& \frac{1}{2} \sum_{\sf m} {\cal B}_{\sf m} 
X^{\sf m}_A,
\nonumber \\ 
{\cal B}^*_{AB} &=& - \sum_{\sf m} {\cal B}_{\sf m}  
X^{\sf m}_{AB}. 
\nonumber 
\end{eqnarray} 
This shows that the angular dependence of these quantities is purely
quadrupolar ($l=2$).  

\subsection{Comoving observer in a spatially-flat cosmology} 

To illustrate how the formalism works we first consider the world line
of a comoving observer in a cosmological spacetime with metric 
\begin{equation} 
ds^2 = -dt^2 + a^2(t)\bigl( dx^2 + dy^2 + dz^2 \bigr), 
\label{2.9.1}
\end{equation}
where $a(t)$ is an arbitrary scale factor; for simplicity we take the
cosmology to be spatially flat. We take the observer to be at the
spatial origin of the coordinate system ($x=y=z=0$), and her velocity
vector is given by 
\begin{equation}  
u^\mu = (1,0,0,0).  
\label{2.9.2}
\end{equation} 
This satisfies the geodesic equation, so $a^\mu = 0$. We wish
to transform the metric of Eq.~(\ref{2.9.1}) to retarded coordinates
based at the world line of this observer.   

To do so we must first construct a triad of orthonormal spatial vectors 
$\base{\mu}{a}$. A simple choice is
\begin{eqnarray} 
\base{\mu}{1} &=& (0,a^{-1},0,0), 
\nonumber \\  
\base{\mu}{2} &=& (0,0,a^{-1},0), 
\label{2.9.3} \\ 
\base{\mu}{3} &=& (0,0,0,a^{-1}); 
\nonumber 
\end{eqnarray}
these vectors are all parallel transported on $\gamma$, and we have
$\omega_{ab} = 0$ according to Eq.~(\ref{2.1.1}). 

Using $\Omega^a = (\sin\theta\cos\phi,\sin\theta\sin\phi,\cos\theta)$
we find that the components of $S_{ab}$, as defined by
Eq.~(\ref{2.4.3}), are given by 
\begin{eqnarray*} 
S_{11} &=& -\ddot{a}/a + (\dot{a}/a)^2 \bigl( \sin^2\theta\sin^2\phi 
+ \cos^2\theta \bigr), \\ 
S_{12} &=& -(\dot{a}/a)^2 \sin^2\theta\sin\phi\cos\phi, \\ 
S_{13} &=& -(\dot{a}/a)^2 \sin\theta\cos\theta\cos\phi, \\ 
S_{22} &=& -\ddot{a}/a + (\dot{a}/a)^2 \bigl( \sin^2\theta\cos^2\phi  
+ \cos^2\theta \bigr), \\ 
S_{23} &=& -(\dot{a}/a)^2 \sin\theta\cos\theta\sin\phi, \\ 
S_{33} &=& -\ddot{a}/a + (\dot{a}/a)^2 \sin^2\theta, 
\end{eqnarray*}
where an overdot indicates differentiation with respect to $t$; the
scale factor and its derivatives are now all evaluated at
$t=u$. According to Eq.~(\ref{2.4.4}), contracting $S_{ab}$ with
$\Omega^b$ gives $S_a$, and we obtain 
\[
S_a = -(\ddot{a}/a) \Omega_a. 
\]
Another contraction with $\Omega^a$ gives 
\[
S = -\ddot{a}/a, 
\]
according to Eq.~(\ref{2.4.5}). From these results it follows that 
\[
S_{ab} \Omega^a_A \Omega^b_B = \bigl[ -\ddot{a}/a + (\dot{a}/a)^2
\bigr] \Omega_{AB},
\]
where $\Omega^a_{A} \equiv \partial \Omega^a/\partial \theta^A$ and
$\Omega_{AB} = \mbox{diag}(1,\sin^2\theta)$ were first introduced in 
Sec.~II G. We also have $S_a \Omega^a_A = 0$. 

Substituting these relations into Eqs.~(\ref{2.7.9})--(\ref{2.7.11})
shows that in the retarded coordinates, the metric components are
given by
\begin{eqnarray} 
g_{uu} &=& -1 + r^2 (\ddot{a}/a) + O(r^3),
\nonumber \\ 
g_{uA} &=& O(r^4), 
\label{2.9.4} \\ 
g_{AB} &=& r^2 \Omega_{AB} \biggl\{ 1 + \frac{1}{3} r^2 \bigl[
\ddot{a}/a - (\dot{a}/a)^2 \bigr] + O(r^3) \biggr\}, 
\nonumber 
\end{eqnarray} 
in addition to $g_{ur} = -1$. Not surprisingly, the metric is
spherically symmetric. Recall that the scale factor and its
derivatives are all functions of retarded time $u$. When the scale
factor behaves as a power law, $a(t) \propto t^\alpha$ with $\alpha$ a
constant, we have $\ddot{a}/a = -\alpha(1-\alpha)/u^2$ and 
$\ddot{a}/a - (\dot{a}/a)^2 = -\alpha/u^2$. When instead the scale
factor behaves as an exponential, $a(t) \propto e^{H t}$ with $H$ a
constant, we have $\ddot{a}/a = H^2$ and $\ddot{a}/a - (\dot{a}/a)^2 
= 0$. 

To help clarify the meaning of these results, we present next an
{\it ab initio} derivation of Eq.~(\ref{2.9.4}). We take the metric of 
Eq.~(\ref{2.9.1}) and switch to conformal time $\eta$, which is
defined by the relation $d\eta = dt/a(t)$. The metric becomes 
\[
ds^2 = a^2(\eta) \bigl( -d\eta^2 + dx^2 + dy^2 + dz^2 \bigr). 
\]
We then introduce spherical coordinates $(\rho,\theta,\phi)$
through the relations $x = \rho\sin\theta\cos\phi$,
$y=\rho\sin\theta\sin\phi$, $z=\rho\cos\theta$, and the null
coordinate $\bar{u} = \eta - \rho$. The metric now reads 
\begin{equation}
ds^2 = a^2(\bar{u} + \rho) \bigl( -d\bar{u}^2 - 2\, d\bar{u}d\rho 
+ \rho^2\, d\Omega^2 \bigr), 
\label{2.9.5}
\end{equation}
where $d\Omega^2 = \Omega_{AB}\, d\theta^A d\theta^B = d\theta^2 +
\sin^2\theta\, d\phi^2$. While $\bar{u}$ is a retarded-time coordinate
and $\rho$ is a radial coordinate, these are distinct from
$u$ and $r$, and Eq.~(\ref{2.9.5}) does not match the form of
Eq.~(\ref{2.9.4}). Since $u$ and $\bar{u}$ can both be used to label
light cones centered at $\rho = 0$, there must exist between them a
relation of the form $u = u(\bar{u})$. And since $r$ and $\rho$ can
both be used to parameterize the null generators of a light cone
$\bar{u} = \mbox{constant}$ (although $\rho$ is not an affine
parameter), there must exist between them a relation of the form
$r=r(\bar{u},\rho)$. We shall now obtain these relations. We recall
that the world line is located at $\rho = r = 0$. 

In the coordinates $(\bar{u},\rho,\theta,\phi)$, the observer's
velocity vector is given by 
\begin{equation} 
u^\mu = \biggl(\frac{1}{a(\bar{u})},0,0,0 \biggr), 
\label{2.9.6}
\end{equation}
where the scale factor is evaluated at $\rho = 0$ and expressed as a
function of $\bar{u}$ only. The null vector 
$\bar{k}_\alpha = -\partial_\alpha \bar{u}$ is tangent to the
null cones $\bar{u} = \mbox{constant}$, and its components are given
by $\bar{k}^\alpha = (0,a^{-2},0,0)$, where the scale factor is now a 
function of $\bar{u}+\rho$. We have 
$\bar{k}_\mu u^\mu = -1/a(\bar{u})$.   

We are seeking a null coordinate $u$ and a null vector field 
$k_\alpha = -\partial_\alpha u$ such that $k_\mu u^\mu = -1$; refer
back to Eqs.~(\ref{2.3.6}) and (\ref{2.3.12}). It is easy to see that
this should be given by $k_\alpha = a(\bar{u}) \bar{k}_\alpha$. We
therefore define $u$ with the statement 
\begin{equation}
du = a(\bar{u})\, d\bar{u}, 
\label{2.9.7}
\end{equation} 
and 
\begin{equation} 
k^\alpha = \biggl( 0, \frac{a(\bar{u})}{a^2(\bar{u}+\rho)}, 0, 0
\biggr) 
\label{2.9.8}
\end{equation} 
is tangent to the light cones $u = \mbox{constant}$. This vector
satisfies the geodesic equation $k^\alpha_{\ ;\beta} k^\beta = 0$, and
the affine parameter on the null generators is $r$. From
Eq.~(\ref{2.9.8}) we have $d\rho/dr = a(\bar{u})/a^2(\bar{u}+\rho)$,
which integrates to 
\begin{equation}
r = \int_0^\rho \frac{a^2(\bar{u}+\rho')}{a(\bar{u})}\, d\rho', 
\label{2.9.9}
\end{equation} 
taking into account the boundary values $r(\bar{u},\rho=0) = 0$. 
Equations (\ref{2.9.7}) and (\ref{2.9.9}) give us the transformation
between the old coordinates $(\bar{u},\rho)$ and the new coordinates
$(u,r)$. 

After applying the coordinate transformation to the metric we obtain 
\begin{equation} 
ds^2 = -\biggl[ \frac{a^2(\bar{u}+\rho)}{a^2(\bar{u})} 
- \frac{2 r_{\bar{u}}}{a(\bar{u})} \biggr]\, du^2 - 2\, dudr 
+ a^2(\bar{u}+\rho)\rho^2\, d\Omega^2, 
\label{2.9.10}
\end{equation}
where $r_{\bar{u}} \equiv \partial r/\partial \bar{u}$. To show that
this matches the results of Eq.~(\ref{2.9.4}), we must evaluate the
integral of Eq.~(\ref{2.9.9}). It is sufficient to work in a
neighborhood of $\rho=0$, and $a^2(\bar{u}+\rho)$ can be expressed as 
a Taylor expansion. This yields 
\[
r = a \rho \biggl[ 1 + \frac{a'}{a} \rho + \frac{1}{3}
  \biggl( \frac{a''}{a} + \frac{a^{\prime 2}}{a^2} \biggr) \rho^2 
+ O(\rho^3) \biggr], 
\]
where a prime indicates differentiation with respect to
$\bar{u}$; here and below, the scale factor and its derivatives are
evaluated at $\rho=0$ and expressed as functions of $\bar{u}$
only. From this we gather that $r_{\bar{u}} = a' \rho + a''
\rho^2 + O(\rho^3)$, and Eq.~(\ref{2.9.10}) gives
\[
g_{uu} = -1 + \biggl( \frac{a''}{a} 
- \frac{a^{\prime 2}}{a^2} \biggr) \rho^2 + O(\rho^3)
\]
and
\[
g_{AB} = a^2 \rho^2 \Omega_{AB} \biggl[ 
1 + 2 \frac{a'}{a} \rho + \biggl( \frac{a''}{a} 
+ \frac{a^{\prime 2}}{a^2} \biggr)\rho^2 + O(\rho)^3 \biggr].   
\]
Expressing these results in terms of $r$ instead of $\rho$, and  
converting $\bar{u}$-derivatives into $u$-derivatives using
Eq.~(\ref{2.9.7}), returns the results of Eq.~(\ref{2.9.4}). 
It should be noted that while Eq.~(\ref{2.9.4}) gives the metric in a
neighborhood of $r=0$, the expression given in Eq.~(\ref{2.9.10})
holds globally.  
 
\subsection{Circular motion in Schwarzschild spacetime}  

As another example we consider the world line of a freely-moving
observer in circular motion around a Schwarzschild black hole. In the
usual Schwarzschild coordinates $(t_s, r_s, \theta_s, \phi_s)$ the
metric is given by 
\begin{eqnarray}
ds^2 &=& -(1-2M/r_s)\, dt^2_s + (1-2M/r_s)^{-1}\, dr^2_s 
\nonumber \\ & & \mbox{} 
+ r^2_s \bigl( d\theta^2_s + \sin^2\theta_s\, d\phi^2_s \bigr),
\label{2.10.1}
\end{eqnarray}
where $M$ is the mass of the black hole. The observer moves on a
circular orbit of radius $r_s = R$ with an angular velocity
$d\phi_s/dt_s = \Omega = \sqrt{M/R^3}$, in the equatorial plane
$\theta_s = \pi/2$. The velocity vector is 
\begin{equation} 
u^\mu = \gamma(1,0,0,\Omega), 
\label{2.10.2}
\end{equation}
where $\gamma = (1-3M/R)^{-1/2}$ is a normalization factor. The motion
is geodesic, and we can once more set $a^\mu = 0$. Because
$R_{\alpha\beta} = 0$ for the Schwarzschild spacetime, we will rely on
the results presented in Sec.~II H. 

The vectors 
\[
\base{\mu}{r} = (0,\beta,0,0), \qquad 
\base{\mu}{\theta} = (0,0,1/R,0), 
\]
and 
\[
\base{\mu}{\phi} = \gamma(\Omega R/\beta,0,0,\beta/R),
\]
where $\beta = (1-2M/R)^{1/2}$, are normalized, mutually orthogonal,
and all orthogonal to $u^\mu$. As such they form a valid set of
spatial vectors, but this choice is not optimal because except for 
$\base{\mu}{\theta}$, the vectors are not parallel transported
on the world line. By forming linear superpositions, however, and
choosing the coefficients appropriately, we can find a set of
parallel-transported vectors $\base{\mu}{a}$. We choose 
$\base{\mu}{1} = \cos(\Omega \tau) \base{\mu}{r} 
- \sin(\Omega \tau) \base{\mu}{\phi}$, $\base{\mu}{2} =
\sin(\Omega \tau) \base{\mu}{r} + \cos(\Omega \tau)
\base{\mu}{\phi}$, and $\base{\mu}{3} =
-\base{\mu}{\theta}$, or 
\begin{eqnarray} 
\base{\mu}{1} &=& \biggl(-\frac{\gamma \Omega R}{\beta}\sin \Phi, 
\beta \cos \Phi, 0, -\frac{\beta\gamma}{R} \sin\Phi \biggr), 
\nonumber \\ 
\base{\mu}{2} &=& \biggl(\frac{\gamma \Omega R}{\beta}\cos \Phi, 
\beta \sin \Phi, 0, \frac{\beta\gamma}{R} \cos\Phi \biggr),
\label{2.10.3} \\ 
\base{\mu}{3} &=& \biggl( 0,0,-\frac{1}{R},0 \biggr), 
\nonumber 
\end{eqnarray}
where $\beta = (1-2M/R)^{1/2}$, $\gamma = (1-3M/R)^{-1/2}$, and  
\begin{equation} 
\Phi = \Omega \tau. 
\label{2.10.4}
\end{equation}
The vectors $\base{\mu}{a}$ are all parallel transported on the world
line, and according to Eq.~(\ref{2.1.1}), we have $\omega_{ab} = 0$.   

The electric part ${\cal E}_{ab}$ of the Riemann tensor is defined by
Eq.~(\ref{2.8.1}), and Eq.~(\ref{2.8.27}) gives its decomposition into
harmonic components ${\cal E}_{\sf m}$. Using the tetrad introduced
previously we find that the nonvanishing components are 
\begin{eqnarray}
{\cal E}_0 &=& \frac{M}{R^2(R-3M)},
\nonumber \\ 
{\cal E}_{2c} &=& -\frac{3M}{2R^3} \frac{R-2M}{R-3M} \cos2\Phi, 
\label{2.10.5} \\ 
{\cal E}_{2s} &=& -\frac{3M}{2R^3} \frac{R-2M}{R-3M} \sin2\Phi. 
\nonumber 
\end{eqnarray} 
The magnetic part ${\cal B}_{ab}$ of the Riemann tensor is defined in
Eq.~(\ref{2.8.1}) and decomposed in harmonic components in
Eq.~(\ref{2.8.28}). Its nonvanishing components are 
\begin{eqnarray} 
{\cal B}_{1c} &=& -\frac{6M\Omega}{R(R-3M)} 
\sqrt{1-\frac{2M}{R}} \cos\Phi, 
\nonumber \\ 
& & \label{2.10.6} \\ 
{\cal B}_{1s} &=& -\frac{6M\Omega}{R(R-3M)} 
\sqrt{1-\frac{2M}{R}} \sin\Phi.  
\nonumber
\end{eqnarray} 
Equivalent results were obtained by Alvi \cite{alvi:00}, based on 
earlier work by Fishbone \cite{fishbone:73} and Marck
\cite{marck:83}. 

To construct the metric we must form the quantities ${\cal E}^*$,
${\cal E}^*_A$, ${\cal E}^*_{AB}$, ${\cal B}^*_A$, and 
${\cal B}^*_{AB}$, defined by Eqs.~(\ref{2.8.5}) and
(\ref{2.8.20})--(\ref{2.8.23}). For this we use Eq.~(\ref{2.8.29}) 
and the spherical harmonics listed in Appendix A. We obtain  
\begin{eqnarray} 
{\cal E}^* &=& \frac{M}{2R^2(R-3M)} \bigl( 3\cos^2\theta - 1) 
\nonumber \\ & & \mbox{} 
- \frac{3M}{2R^3} \frac{R-2M}{R-3M} \sin^2\theta \cos2(\phi-\Phi), 
\nonumber \\ 
{\cal E}^*_\theta &=& -\frac{3M}{2R^3(R-3M)} \bigl[ R 
\nonumber \\ & & \mbox{} 
+ (R-2M) \cos2(\phi-\Phi) \bigr] \sin\theta \cos\theta, 
\nonumber \\ 
{\cal E}^*_\phi &=& \frac{3M}{2R^3} \frac{R-2M}{R-3M} \sin^2\theta
\sin2(\phi-\Phi), 
\nonumber \\
{\cal E}^*_{\theta\theta} &=& \frac{3M}{2R^2(R-3M)} \sin^2\theta 
\nonumber \\ & & \mbox{} 
- \frac{3M}{2R^3} \frac{R-2M}{R-3M} \bigl( 1 + \cos^2\theta \bigr) 
\cos2(\phi-\Phi), 
\nonumber \\ 
{\cal E}^*_{\theta\phi} &=& \frac{3M}{R^3} \frac{R-2M}{R-3M}
\sin\theta \cos\theta \sin2(\phi-\Phi), 
\nonumber \\ 
{\cal E}^*_{\phi\phi} &=& -\frac{3M}{2R^2(R-3M)} \sin^4\theta 
\nonumber \\ & & \mbox{} 
+ \frac{3M}{2R^3} \frac{R-2M}{R-3M} \sin^2\theta 
\bigl( 1 + \cos^2\theta \bigr) \cos2(\phi-\Phi),
\nonumber \\
& & \label{2.10.7}
\end{eqnarray} 
and 
\begin{eqnarray} 
{\cal B}^*_\theta &=& -\frac{3M\Omega}{R(R-3M)} 
\sqrt{1-\frac{2M}{R}} \cos\theta \sin(\phi-\Phi), 
\nonumber \\ 
{\cal B}^*_\phi &=& \frac{3M\Omega}{R(R-3M)} \sqrt{1-\frac{2M}{R}} 
\nonumber \\ & & \mbox{} \times 
\sin\theta \bigl(1-2\cos^2\theta) \cos(\phi-\Phi), 
\nonumber \\ 
{\cal B}^*_{\theta\theta} &=& \frac{6M\Omega}{R(R-3M)} 
\sqrt{1-\frac{2M}{R}} \sin\theta \sin(\phi-\Phi),
\nonumber \\ 
{\cal B}^*_{\theta\phi} &=& \frac{6M\Omega}{R(R-3M)} 
\sqrt{1-\frac{2M}{R}} \sin^2\theta\cos\theta \cos(\phi-\Phi),
\nonumber \\ 
{\cal B}^*_{\phi\phi} &=& -\frac{6M\Omega}{R(R-3M)} 
\sqrt{1-\frac{2M}{R}} \sin^3\theta \sin(\phi-\Phi). 
\nonumber \\
& & \label{2.10.8} 
\end{eqnarray}
In these expressions, the components of the Riemann tensor are all
evaluated at $\tau = u$, so that $\Phi = \Omega u$. Substituting them
into Eqs.~(\ref{2.8.16})--(\ref{2.8.19}) gives the Schwarzschild
metric in the retarded coordinates $(u,r,\theta,\phi)$; these are
based at the world line of an observer moving on a circular orbit of
radius $R$ with an angular velocity $\Omega = \sqrt{M/R^3}$.    

\section{Motion of a small black hole in an external universe} 
    
\subsection{Matched asymptotic expansions} 

In this section we consider a nonrotating black hole of small mass $m$
moving in a background spacetime with metric
$g_{\alpha\beta}$, and we seek to determine the equations that govern 
its motion. We will employ the powerful technique of {\it matched
asymptotic expansions} \cite{thorne-hartle:85, manasse:63, kates:80,
death:96, mino-etal:97, alvi:00, detweiler:01} and make use of the
retarded coordinates developed in Sec.~II.   
   
The problem presents itself with a clean separation of length scales, 
and the method relies heavily on this. On the one hand we have the 
length scale associated with the small black hole, which is set by its
mass $m$. On the other hand we have the length scale associated with 
the background spacetime, which is set by the radius of curvature 
$\cal R$; this is defined so that a typical component of the
background spacetime's Riemann tensor is equal to $1/{\cal R}^2$ up to
a numerical factor of order unity. We demand that 
$m/{\cal R} \ll 1$. For simplicity we assume that the background
spacetime contains no matter, so that its metric is a solution to the
Einstein field equations in vacuum. 

Let $r$ be a meaningful measure of distance from the small black hole,  
and let us consider a region of spacetime defined by $r < r_i$, where
$r_i$ is a constant that is much smaller than ${\cal R}$. This
inequality defines a narrow world tube that surrounds the black 
hole, and we shall call this region the {\it internal zone}. In the
internal zone the gravitational field is dominated by the black hole,
and the metric can be expressed as     
\begin{equation}
{\sf g}(\mbox{internal zone}) = g(\mbox{black hole}) 
+ H_1/{\cal R} + H_2/{\cal R}^2 + \cdots, 
\label{3.1.1}
\end{equation}
where $g(\mbox{black hole})$ is the metric of a nonrotating black hole  
in isolation (as given by the unperturbed Schwarzschild solution),
while $H_1$ and $H_2$ are corrections associated with the
conditions in the external universe. The metric of Eq.~(\ref{3.1.1})
represents a black hole that is distorted by the tidal gravitational
field of the external universe, and $H_1$, $H_2$ are functions of 
the spacetime coordinates that can be obtained by solving the
Einstein field equations. They must be such that the spacetime 
possesses a regular event horizon near $r=2m$, and such that
${\sf g}(\mbox{internal zone})$ agrees with the metric of the external 
universe --- the metric of the background spacetime in the absence of
a black hole --- when $r \gg m$. As we shall see in Sec.~III B,
$H_1$ actually vanishes and the small correction $H_2/{\cal R}^2$ can
be obtained by employing the well-developed tools of black-hole
perturbation theory \cite{regge-wheeler:57, zerilli:70,
vishveshwara:70, detweiler:01, jhingan-tanaka:03}.         

Consider now a region of spacetime defined by $r > r_e$, where $r_e$  
is a constant that is much larger than $m$; this region will be called 
the {\it external zone}. In the external zone the gravitational field
is dominated by the conditions in the external universe, and the
metric can be expressed as  
\begin{eqnarray}
{\sf g}(\mbox{external zone}) &=& g(\mbox{background spacetime})
\nonumber \\ & & \mbox{}  
+ m h_1 + m^2 h_2 + \cdots, 
\label{3.1.2}
\end{eqnarray}
where $g(\mbox{background spacetime})$ is the unperturbed metric of 
the background spacetime in which the black hole is moving, while
$h_1$ and $h_2$ are corrections associated with the hole's
presence; these are functions of the spacetime
coordinates that can be obtained by solving the Einstein
field equations. We shall truncate Eq.~(\ref{3.1.2}) to first order 
in $m$, and $m h_1$ will be calculated in Sec.~III C by linearizing
the field equations about the metric of the background spacetime. 

The metric ${\sf g}(\mbox{external zone})$ is a functional of a world
line $\gamma$ that represents the motion of the small black hole in
the background spacetime. Our goal is to obtain a description of this 
world line, in the form of equations of motion to be satisfied by the
black hole; these equations will be formulated in the background
spacetime.  It is important to understand that fundamentally, $\gamma$ 
exists only as an external-zone construct: It is only in the external
zone that the black hole can be thought of as moving on a world line;
in the internal zone the black hole is revealed as an extended object
and the notion of a world line describing its motion is no longer 
meaningful.  

Equations (\ref{3.1.1}) and (\ref{3.1.2}) give two different 
expressions for the metric of the same spacetime; the first is valid
in the internal zone $r < r_i \ll {\cal R}$, while the second is valid 
in the external zone $r > r_e \gg m$. The fact that ${\cal R} \gg m$
allows us to define a {\it buffer zone} in which $r$ is restricted to
the interval $r_e < r < r_i$. In the buffer zone $r$ is simultaneously
much larger than $m$ and much smaller than ${\cal R}$ --- a typical
value might be $\sqrt{m {\cal R}}$ --- and Eqs.~(\ref{3.1.1}),
(\ref{3.1.2}) are simultaneously valid. Since the two metrics are the
same up to a diffeomorphism, these expressions must agree. And since 
${\sf g}(\mbox{external zone})$ is a functional of a world line
$\gamma$ while ${\sf g}(\mbox{internal zone})$ contains no such
information, {\it matching the metrics necessarily determines the
motion of the small black hole in the background spacetime}. 

Matching the metrics of Eqs.~(\ref{3.1.1}) and (\ref{3.1.2}) in the 
buffer zone can be carried out in practice only after performing a 
transformation from the external coordinates used to express 
${\sf g}(\mbox{external zone})$ to the internal coordinates employed
for ${\sf g}(\mbox{internal zone})$. The details of this coordinate
transformation are presented in Appendix C and the end result of 
matching is revealed in Sec.~III D.  

\subsection{Metric in the internal zone} 

To proceed with the program outlined in the previous subsection we 
first calculate the internal-zone metric and replace
Eq.~(\ref{3.1.1}) by a more concrete expression. We recall that the 
internal zone is defined by $r < r_i \ll {\cal R}$, where $r$ is a
suitable measure of distance from the black hole.  

We begin by expressing $g(\mbox{black hole})$, the Schwarzschild
metric of an isolated black hole of mass $m$, in terms of retarded
Eddington-Finkelstein coordinates $(\bar{u},\bar{r},\bar{\theta}^A)$,
where $\bar{u}$ is retarded time, $\bar{r}$ the usual areal radius,
and $\bar{\theta}^A = (\bar{\theta},\bar{\phi})$ are two angles
on the spheres of constant $\bar{u}$ and $\bar{r}$. The metric is
given by  
\begin{equation}
ds^2 = - f\, d\bar{u}^2 - 2\, d\bar{u} d\bar{r} + \bar{r}^2\,
d\bar{\Omega}^2, 
\label{3.2.1}
\end{equation} 
where
\begin{equation} 
f = 1 - \frac{2m}{\bar{r}}, 
\label{3.2.2}
\end{equation}
and $d\bar{\Omega}^2 = \bar{\Omega}_{AB}\, d\bar{\theta}^A
d\bar{\theta}^B = d\bar{\theta}^2 + \sin^2\bar{\theta}\,
d\bar{\phi}^2$ is the line element on the unit two-sphere. In the 
limit $r \gg m$ this metric achieves the asymptotic values 
$g_{\bar{u}\bar{u}} \to -1$, $g_{\bar{u}\bar{r}} = -1$, 
$g_{\bar{u}\bar{A}} = 0$, and $g_{\bar{A}\bar{B}} = \bar{r}^2\,
\bar{\Omega}_{AB}$; these are appropriate for a black hole immersed in
a flat spacetime charted by retarded coordinates.  

The corrections $H_1$ and $H_2$ in Eq.~(\ref{3.1.1})  
encode the information that our black hole is not isolated but
in fact immersed in an external universe whose metric becomes 
$g(\mbox{background spacetime})$ asymptotically. In the internal zone
the metric of the background spacetime can be expanded in powers of
$\bar{r}/{\cal R}$ and expressed in a form that can be directly
imported from Sec.~II. If we assume that the
``world line'' $\bar{r} = 0$ has no acceleration in the background 
spacetime (a statement that will be justified shortly), then the   
asymptotic values of ${\sf g}(\mbox{internal zone})$
must be given by Eqs.~(\ref{2.8.16})--(\ref{2.8.19}):   
\begin{eqnarray*}  
{\sf g}_{\bar{u}\bar{u}} &\to& - 1 - \bar{r}^2 \bar{{\cal E}}^*  
+ O(\bar{r}^3/{\cal R}^3), \\ 
{\sf g}_{\bar{u}\bar{r}} &=& - 1, \\ 
{\sf g}_{\bar{u}\bar{A}} &\to& 
\frac{2}{3} \bar{r}^3 \bigl( \bar{{\cal E}}^*_A   
+ \bar{{\cal B}}^*_A \bigr) + O(\bar{r}^4/{\cal R}^3), \\  
{\sf g}_{\bar{A}\bar{B}} &\to& \bar{r}^2 \bar{\Omega}_{AB}  
- \frac{1}{3} \bar{r}^4 \bigl( \bar{{\cal E}}^*_{AB} 
+ \bar{{\cal B}}^*_{AB} \bigr) + O(\bar{r}^5/{\cal R}^3),    
\end{eqnarray*} 
where 
\begin{eqnarray*} 
\bar{{\cal E}}^* &=& {\cal E}_{ab} \bar{\Omega}^a \bar{\Omega}^b, \\   
\bar{{\cal E}}^*_A &=& {\cal E}_{ab} \bar{\Omega}^a_A \bar{\Omega}^b, \\ 
\bar{{\cal E}}^*_{AB} &=& 2 {\cal E}_{ab} \bar{\Omega}^a_A
\bar{\Omega}^b_B + \bar{{\cal E}}^* \bar{\Omega}_{AB}  
\end{eqnarray*}
and
\begin{eqnarray*}  
\bar{{\cal B}}^*_A &=& \varepsilon_{abc} \bar{\Omega}^a_A
\bar{\Omega}^b {\cal B}^c_{\ d} \bar{\Omega}^d, \\ 
\bar{{\cal B}}^*_{AB} &=& 2 \varepsilon_{acd} \bar{\Omega}^c  
{\cal B}^d_{\ b} \bar{\Omega}^{(a}_A \bar{\Omega}^{b)}_B  
\end{eqnarray*}  
are the tidal gravitational fields that were first introduced in
Sec.~II H. Recall that $\bar{\Omega}^a =
(\sin\bar{\theta}\cos\bar{\phi}, \sin\bar{\theta}\sin\bar{\phi}, 
\cos\bar{\theta})$ and $\bar{\Omega}^a_A = 
\partial \bar{\Omega}^a/\partial \bar{\theta}^A$. Apart from an 
angular dependence made explicit by these relations, the tidal fields 
depend on $\bar{u}$ through the frame components ${\cal E}_{ab} \equiv 
R_{a0b0} = O(1/{\cal R}^2)$ and ${\cal B}^a_{\ b} \equiv \frac{1}{2}
\varepsilon^{acd} R_{0bcd} = O(1/{\cal R}^2)$ of the Riemann tensor. 
(This is the Riemann tensor of the background spacetime evaluated at 
$\bar{r}=0$.) Notice that we have incorporated the fact that the Ricci
tensor vanishes at $\bar{r}=0$: the black hole moves in a vacuum
spacetime.      

The modified asymptotic values lead us to the following ansatz for the 
internal-zone metric: 
\begin{eqnarray} 
{\sf g}_{\bar{u}\bar{u}} &=& - f \bigr[ 1 + \bar{r}^2 e_1(\bar{r})  
\bar{\cal E}^* \bigr] + O(\bar{r}^3/{\cal R}^3), 
\label{3.2.3} \\ 
{\sf g}_{\bar{u}\bar{r}} &=& - 1,
\label{3.2.4} \\  
{\sf g}_{\bar{u}\bar{A}} &=& \frac{2}{3} \bar{r}^3 \bigl[ e_2(\bar{r})   
\bar{\cal E}^*_A + b_2(\bar{r}) \bar{\cal B}^*_A \bigr] 
+ O(\bar{r}^4/{\cal R}^3),
\label{3.2.5} \\   
{\sf g}_{\bar{A}\bar{B}} &=& \bar{r}^2 \bar{\Omega}_{AB} 
- \frac{1}{3} \bar{r}^4    
\big[ e_3(\bar{r}) \bar{\cal E}^*_{AB} + b_3(\bar{r}) 
\bar{\cal B}^*_{AB} \bigr] 
\nonumber \\ & & \mbox{} 
+ O(\bar{r}^5/{\cal R}^3).
\label{3.2.6}
\end{eqnarray}
The five functions $e_1$, $e_2$, $e_3$, $b_2$, and $b_3$ can 
all be determined by solving the Einstein field equations; they must
approach unity when $r \gg m$ and be well-behaved at $r=2m$
(so that the tidally distorted black hole will have a nonsingular
event horizon). In Appendix B, I show that they are given by 
\begin{equation} 
e_1(\bar{r}) = e_2(\bar{r}) = f, \qquad  
e_3(\bar{r}) = 1 - \frac{2m^2}{\bar{r}^2}, 
\label{3.2.7}
\end{equation} 
and 
\begin{equation} 
b_2(\bar{r}) = f, \qquad  
b_3(\bar{r}) = 1. 
\label{3.2.8}
\end{equation} 
It is clear from Eqs.~(\ref{3.2.3})--(\ref{3.2.6}) that the assumed
deviation of ${\sf g}(\mbox{internal zone})$ with respect to 
$g(\mbox{black hole})$ scales as $1/{\cal R}^2$. It is therefore of
the form of Eq.~(\ref{3.1.1}) with $H_1 = 0$. The fact that $H_1$
vanishes comes as a consequence of our previous assumption that the
``world line'' $\bar{r} = 0$ has a zero acceleration in the background
spacetime; a nonzero acceleration of order $1/{\cal R}$ would bring
terms of order $1/{\cal R}$ to the metric, and $H_1$ would then be
nonzero. The perturbed metric of Eqs.~(\ref{3.2.3})--(\ref{3.2.8})
differs from the one presented by Detweiler \cite{detweiler:01} only
by a transformation from Schwarzschild to Eddington-Finkelstein
coordinates, and a transformation from the Zerilli gauge
\cite{zerilli:70} gauge adopted by him to the retarded gauge adopted
here.        

Why is the assumption of no acceleration justified? As I shall
explain more fully in the next paragraph, the reason is simply that it 
reflects a choice of coordinate system: setting the acceleration to
zero amounts to adopting a specific --- and convenient --- gauge
condition.        

Inspection of Eqs.~(\ref{3.2.3})--(\ref{3.2.6}) reveals that the 
angular dependence of the metric perturbation is generated entirely by  
scalar, vectorial, and tensorial spherical harmonics of degree
$l=2$; this observation was elaborated toward the end of Sec.~II H and
in Appendix A. In particular, $H_2$ contains no $l=0$ and $l=1$ modes,
and this statement reflects a choice of gauge condition. Zerilli has
shown \cite{zerilli:70} that a perturbation of the Schwarzschild
spacetime with $l = 0$ corresponds to a shift in the mass
parameter. As Thorne and Hartle have shown \cite{thorne-hartle:85}, a 
black hole interacting with its environment will undergo a change of
mass, but this effect is of order $m^3/{\cal R}^2$ and thus beyond the
level of accuracy of our calculations. There is therefore no need to
include $l = 0$ terms in $H_2$. Similarly, it was shown by Zerilli
that odd-parity perturbations of degree $l=1$ correspond to a shift
in the black hole's angular-momentum parameters. As Thorne and Hartle
have shown, a change of angular momentum is quadratic in the hole's
angular momentum, and we can ignore this effect when dealing with a
nonrotating black hole. There is therefore no need to include
odd-parity, $l=1$ terms in $H_2$. Finally, Zerilli has shown
that in a vacuum spacetime, even-parity perturbations of degree
$l=1$ correspond to a change of coordinate system --- these
modes are pure gauge. Since we have the freedom to adopt any gauge 
condition, we can exclude even-parity, $l=1$ terms from the perturbed
metric. This leads us to Eqs.~(\ref{3.2.3})--(\ref{3.2.6}), which
contain only $l=2$ perturbation modes; the even-parity modes are
contained in those terms that involve ${\cal E}_{ab}$, while the
odd-parity modes are associated with ${\cal B}_{ab}$. The perturbed
metric contains also higher multipoles, but those come at a higher
order in $1/{\cal R}$; for example, the terms of order $1/{\cal R}^3$
include $l=3$ modes. We conclude that
Eqs.~(\ref{3.2.3})--(\ref{3.2.6}) is a sufficiently general ansatz for
the metric in the internal zone.   

It shall prove convenient to transform 
${\sf g}(\mbox{internal zone})$ from the quasi-spherical coordinates
$(\bar{r},\bar{\theta}^A)$ to a set of quasi-Cartesian coordinates
$\bar{x}^a = \bar{r} \bar{\Omega}^a(\bar{\theta}^A)$; the
transformation rules are worked out in Sec.~II G. This gives  
\begin{eqnarray} 
{\sf g}_{\bar{u}\bar{u}} &=& -f \bigl(1 + \bar{r}^2 f \bar{\cal E}^*  
\bigr) + O(\bar{r}^3/{\cal R}^3), 
\label{3.2.9} \\ 
{\sf g}_{\bar{u}\bar{a}} &=& -\bar{\Omega}_a + \frac{2}{3} \bar{r}^2 
f \bigl( \bar{\cal E}^*_a + \bar{\cal B}^*_a \bigr) 
+ O(\bar{r}^3/{\cal R}^3),  
\label{3.2.10} \\ 
{\sf g}_{\bar{a}\bar{b}} &=& \delta_{ab} 
- \bar{\Omega}_a \bar{\Omega}_b  
- \frac{1}{3} \bar{r}^2 \biggl( 1 - 2\frac{m^2}{\bar{r}^2} \biggr)  
\bar{\cal E}^*_{ab} - \frac{1}{3} \bar{r}^2 \bar{\cal B}^*_{ab} 
\nonumber \\ & & \mbox{} 
+ O(\bar{r}^3/{\cal R}^3), 
\label{3.2.11}
\end{eqnarray} 
where $f = 1 -2m/\bar{r}$ and where the tidal fields 
\begin{eqnarray*}  
\bar{\cal E}^* &=& {\cal E}_{ab} \bar{\Omega}^a \bar{\Omega}^b, \\ 
\bar{\cal E}^*_a &=& \bigl(\delta_a^{\ b} - \bar{\Omega}_a
\bar{\Omega}^b \bigr) {\cal E}_{bc} \bar{\Omega}^c, \\ 
\bar{\cal E}^*_{ab} &=& 2 {\cal E}_{ab} 
- 2\bar{\Omega}_a {\cal E}_{bc} \bar{\Omega}^c  
- 2\bar{\Omega}_b {\cal E}_{ac} \bar{\Omega}^c
+ (\delta_{ab} + \bar{\Omega}_a \bar{\Omega}_b) \bar{\cal E}^*, \\ 
\bar{\cal B}^*_a &=& \varepsilon_{abc} \bar{\Omega}^b {\cal B}^c_{\ d}
\bar{\Omega}^d, \\ 
\bar{\cal B}^*_{ab} &=& \varepsilon_{acd} \bar{\Omega}^c 
{\cal B}^d_{\ e} \bigl(\delta^e_{\ b} - \bar{\Omega}^e \bar{\Omega}_b
\bigr) + \varepsilon_{bcd} \bar{\Omega}^c {\cal B}^d_{\ e} 
\bigl(\delta^e_{\ a} - \bar{\Omega}^e \bar{\Omega}_a \bigr), 
\end{eqnarray*}    
were first introduced in Sec.~II H. The metric of
Eqs.~(\ref{3.2.9})--(\ref{3.2.11}) represents the spacetime
geometry of a black hole immersed in an external universe and 
distorted by its tidal gravitational field.   

\subsection{Metric in the external zone} 

We next move on to the external zone and seek to replace
Eq.~(\ref{3.1.2}) by a more concrete expression; recall that the  
external zone is defined by $m \ll r_e < r$. We take advantage of the
fact that in the external zone, the gravitational perturbation
associated with the presence of a black hole cannot be distinguished
from the perturbation produced by a point particle of the same mass. 

The external-zone metric is decomposed as 
\begin{equation}
{\sf g}_{\alpha\beta} = g_{\alpha\beta} + h_{\alpha\beta}, 
\label{3.3.1}
\end{equation}
where $g_{\alpha\beta}$ is the metric of the background spacetime and  
$h_{\alpha\beta} = O(m)$ is the perturbation; we shall work
consistently to first order in $m$ and systematically discard all
terms of higher order. We relate $h_{\alpha\beta}$ to trace-reversed
potentials $\gamma_{\alpha\beta}$, 
\begin{equation}
h_{\alpha\beta} = \gamma_{\alpha\beta} - \frac{1}{2} \bigl(
g^{\gamma\delta} \gamma_{\gamma\delta} \bigr) g_{\alpha\beta}, 
\label{3.3.2}
\end{equation}
and we impose the Lorenz gauge condition 
\begin{equation} 
\gamma^{\alpha\beta}_{\ \ \ ;\beta} = 0; 
\label{3.3.3}
\end{equation}
indices are raised and lowered with $g^{\alpha\beta}$ and
$g_{\alpha\beta}$, respectively. With the understanding that the
background spacetime contains no matter, linearizing the Einstein
field equations produces the wave equation 
\begin{equation} 
\Box \gamma^{\alpha\beta} + 2 R^{\alpha\ \beta}_{\ \gamma\ \delta} 
\gamma^{\gamma\delta} = -16\pi T^{\alpha\beta}
\label{3.3.4}
\end{equation}
for the potentials. Here, $\Box = g^{\alpha\beta}
\nabla_\alpha \nabla_\beta$ is the wave operator and 
\begin{equation} 
T^{\alpha\beta}(x) = m \int_\gamma g^\alpha_{\ \mu}(x,z)  
g^\beta_{\ \nu}(x,z) u^\mu u^\nu \delta_4(x,z)\, d\tau 
\label{3.3.5}
\end{equation} 
is the stress-energy tensor of a point particle of mass $m$ traveling
on a world line $\gamma$; $\delta_4(x,z)$ is a scalarized, 
four-dimensional Dirac functional, the world line is described by
relations $z^\mu(\tau)$ in which $\tau$ is proper time, and $u^\mu = 
dz^\mu/d\tau$ is the particle's velocity vector. Solving the
linearized field equations produces   
\begin{equation} 
\gamma_{\alpha\beta}(x) = 4 m \int_\gamma 
G_{\alpha\beta\mu\nu}(x,z) u^\mu u^\nu\, d\tau,    
\label{3.3.6}
\end{equation}
where $G_{\alpha\beta\mu\nu}(x,z)$ is the retarded Green's
function \footnote{The normalization of the gravitational Green's 
function varies from author to author. Here the normalization is such
that the Green's function obeys a wave equation with a right-hand side
given by $-4\pi g^{(\alpha}_{\ \mu}(x,z) g^{\beta)}_{\ \nu}(x,z)
\delta_4(x,z)$. The factor of $4\pi$ accounts for the factor of 4 on
the right-hand side of Eq.~(\ref{3.3.6}).} \cite{sciama-etal:69}
associated with Eq.~(\ref{3.3.4}).  

We now place ourselves in the buffer zone (where $m \ll r \ll  
{\cal R}$ and where the matching will take place) and work toward
expressing ${\sf g}(\mbox{external zone})$ as an expansion in powers 
of $r/{\cal R}$. For this purpose we adopt the retarded
coordinates $(u,r\Omega^a)$ of Sec.~II and rely on the machinery 
developed there.   

We begin with $g_{\alpha\beta}$, the metric of the background 
spacetime. We have seen in Sec.~II H that if the world line
$\gamma$ is a geodesic, if the vectors $\base{\mu}{a}$ are 
parallel transported on the world line, and if the Ricci tensor
vanishes on $\gamma$, then the metric takes the form given by 
Eqs.~(\ref{2.8.13})--(\ref{2.8.15}). This form, however, is too
restrictive for our purposes: We must allow $\gamma$ to have an 
acceleration, and allow the basis vectors to be transported in the 
most general way compatible with their orthonormality property; this 
transport law is given by Eq.~(\ref{2.1.1}),   
\[
\frac{D \base{\mu}{a}}{d \tau} = a_a u^{\mu} 
+ \omega_a^{\ b} \base{\mu}{b},  
\] 
where $a_a(\tau) = a_{\mu} \base{\mu}{a}$ are the frame components of 
the acceleration vector $a^\mu = D u^\mu/d\tau$, and
$\omega_{ab}(\tau) = -\omega_{ba}(\tau)$ is a rotation tensor to be  
determined. Anticipating that $a_a$ and $\omega_{ab}$ will both be
proportional to $m$, we express the metric of the background spacetime 
as  
\begin{eqnarray}  
g_{uu} &=& - 1 - 2 r a_a \Omega^a - r^2 {\cal E}^* 
+ O(r^3/{\cal R}^3), 
\label{3.3.7} \\ 
g_{ua} &=& -\Omega_a + r \bigr(\delta_a^{\ b} - \Omega_a
\Omega^b \bigr) a_b - r \omega_{ab} \Omega^b 
\nonumber \\ & & \mbox{} 
+ \frac{2}{3} r^2 \bigl( {\cal E}^*_a  + {\cal B}^*_a \bigr) 
+ O(r^3/{\cal R}^3),  
\label{3.3.8} \\ 
g_{ab} &=& \delta_{ab} - \Omega_a \Omega_b - \frac{1}{3} r^2  
\bigl( {\cal E}^*_{ab} + {\cal B}^*_{ab} \bigr) 
\nonumber \\ & & \mbox{} 
+ O(r^3/{\cal R}^3),   
\label{3.3.9}
\end{eqnarray}   
where ${\cal E}^*$, ${\cal E}^*_a$, ${\cal E}^*_{ab}$, ${\cal B}^*_a$,
and ${\cal B}^*_{ab}$ are the tidal gravitational fields first
introduced in Sec.~II H. The metric of
Eqs.~(\ref{3.3.7})--(\ref{3.3.9}) is obtained from the general form
of Eqs.~(\ref{2.6.3})--(\ref{2.6.5}) by neglecting terms quadratic in   
$a_a$ and $\omega_{ab}$ and specializing to a zero Ricci tensor. 

\begin{figure}
\includegraphics[angle=0,scale=0.4]{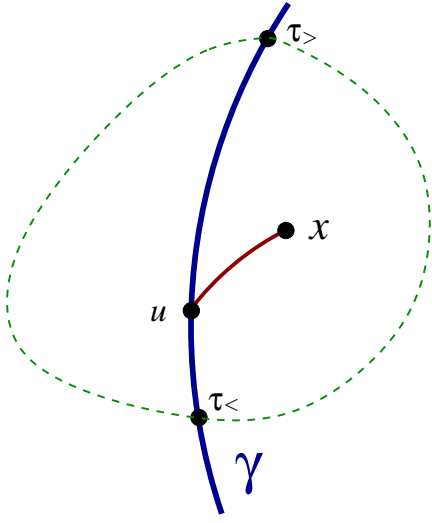} 
\caption{The region within the dashed boundary represents the normal  
convex neighborhood of the point $x$. The world line $\gamma$ enters  
the neighborhood at proper time $\tau_<$ and exits at proper time
$\tau_>$. Also shown is the retarded point $x' \equiv z(u)$ and the
null geodesics that links $x$ to the world line.} 
\end{figure} 

To express the perturbation $h_{\alpha\beta}$ as an expansion in
powers of $r/{\cal R}$ we assume that $x$ is sufficiently close to
$\gamma$ that a portion of the world line traverses ${\cal N}(x)$,
the normal convex neighborhood of the point $x$ (see Fig.~2); this
assumption is compatible with the condition $r \ll {\cal R}$. We then   
re-express Eq.~(\ref{3.3.6}) as 
\[
\gamma_{\alpha\beta} = 4m \biggl( 
\int_{-\infty}^{\tau_<} + \int_{\tau_<}^{\tau_>} 
+ \int_{\tau_>}^\infty \biggr) 
G_{\alpha\beta\mu\nu} u^\mu u^\nu\, d\tau, 
\]
where $\tau_<$ and $\tau_>$ are the values of the proper-time
parameter at which $\gamma$ enters and leaves ${\cal N}(x)$,
respectively. The third integration contributes nothing because $x$ is
then in the past of $z(\tau)$ and the retarded Green's function
vanishes. For the second integration, $x$ is the normal convex
neighborhood of $z(\tau)$, and the retarded Green's function can be
put in the Hadamard form \cite{hadamard:23,
dewitt-brehme:60, sciama-etal:69, friedlander:75}  
\begin{equation}
G_{\alpha\beta\mu\nu}(x,z) = U_{\alpha\beta\mu\nu}(x,z)
\delta_+(\sigma) + V_{\alpha\beta\mu\nu}(x,z) \theta_+(-\sigma), 
\label{3.3.10}
\end{equation}
where $U_{\alpha\beta\mu\nu}(x,z)$ and $V_{\alpha\beta\mu\nu}(x,z)$
are smooth bitensors \footnote{The tail part of the Green's function
is denoted $V_{\alpha\beta\mu\nu}(x,z)$ in this work, but most authors
insert a minus sign and call it $-v_{\alpha\beta\mu\nu}(x,z)$.},
$\sigma(x,z)$ is Synge's world function \cite{synge:60,
dewitt-brehme:60}, and $\delta_+(\sigma)$ and $\theta_+(\sigma)$ are
Dirac and Heaviside distributions restricted to the future of
$z(\tau)$ \cite{dewitt-brehme:60, friedlander:75}. To integrate over
the Dirac term we change variables from $\tau$ to $\sigma$, noticing 
that $\sigma$ increases as $z(\tau)$ passes through the retarded point
$z(u) \equiv x'$; recalling the definition of retarded distance given
in Eq.~(\ref{2.2.2}), the integral evaluates to
$U_{\alpha\beta\gamma'\delta'}(x,x') u^{\gamma'}u^{\delta'}/r$. The
integration over the Heaviside term is cut off at $\tau = u$, and we
obtain our final expression for the perturbation: 
\begin{equation}
\gamma_{\alpha\beta}(x) = \frac{4m}{r} 
U_{\alpha\beta\gamma'\delta'}(x,x') u^{\gamma'} u^{\delta'}  
+ \gamma^{\rm tail}_{\alpha\beta}(x). 
\label{3.3.11}
\end{equation}
Here, primed indices refer to the retarded point $x' \equiv z(u)$    
associated with $x$, and 
\begin{eqnarray} 
\gamma^{\rm tail}_{\alpha\beta}(x) &=& 
4 m \int_{\tau_<}^u V_{\alpha\beta\mu\nu}(x,z) u^\mu u^\nu\,
d\tau 
\nonumber \\ & & \mbox{} 
+ 4 m \int_{-\infty}^{\tau_<} 
G_{\alpha\beta\mu\nu}(x,z) u^\mu u^\nu\, d\tau 
\nonumber \\ 
&\equiv& 4 m \int_{-\infty}^{u^-} 
G_{\alpha\beta\mu\nu}(x,z) u^\mu u^\nu\, d\tau 
\label{3.3.12}
\end{eqnarray}
is the ``tail part'' of the gravitational potentials. Notice that I 
have introduced a short-hand notation in the last line of
Eq.~(\ref{3.3.12}); the important point is that the integral of 
$G_{\alpha\beta\mu\nu}(x,z) u^\mu u^\nu$ up to $\tau = u^-$ avoids the 
singular behavior of the Green's function on the light cone
$\sigma(x,z) = 0$.     

We must next express Eq.~(\ref{3.3.11}) in the form of an expansion in 
powers of $r/{\cal R}$. For this we shall need the expansion
\footnote{This result is implicitly contained in Appendix A of 
Ref.~\cite{mino-etal:97}. It is derived from scratch in Sec.~15 of
Ref.~\cite{poisson:03}.} 
\[ 
U_{\alpha\beta\gamma'\delta'} u^{\gamma'} u^{\delta'} =  
g^{\alpha'}_{\ (\alpha} g^{\beta'}_{\ \beta)} \Bigl[ u_{\alpha'}
u_{\beta'} + O(r^3/{\cal R}^3) \Bigr], 
\]
which contains no terms at order $r/{\cal R}$ or $(r/{\cal R})^2$. We
shall also need the general expansion of a tensor $A_{\alpha\beta}(x)$
in terms of its values at a neighboring point $x'$ \footnote{This is
Eq.~(A21) of Ref.~\cite{mino-etal:97}. This result is derived from
scratch in Sec.~5 of Ref.~\cite{poisson:03}.}: 
\[
A_{\alpha\beta}(x) = g^{\alpha'}_{\ \alpha} g^{\beta'}_{\ \beta}   
\biggl( A_{\alpha'\beta'} 
- A_{\alpha'\beta';\gamma'}\, \sigma^{\gamma'} + \cdots \biggr). 
\]
Here $A_{\alpha\beta}$ will stand for $\gamma^{\rm
tail}_{\alpha\beta}$ and $x'$ will be the retarded point $z(u)$
associated with $x$; accordingly, Eq.~(\ref{2.2.3}) gives 
$\sigma^{\alpha'} = -r (u^{\alpha'} + \Omega^a \base{\alpha'}{a})$. 
Combining all these results, Eq.~(\ref{3.3.11}) becomes 
\begin{eqnarray} 
\gamma_{\alpha\beta}(x) &=& 
g^{\alpha'}_{\ \alpha} g^{\beta'}_{\ \beta}  
\biggl[ \frac{4m}{r} u_{\alpha'} u_{\beta'} 
+ \gamma^{\rm tail}_{\alpha'\beta'}  
\nonumber \\ & & \mbox{} 
+ r \gamma^{\rm tail}_{\alpha'\beta'\gamma'} \bigl( u^{\gamma'} 
+ \Omega^c \base{\gamma'}{c} \bigr) 
\nonumber \\ & & \mbox{} 
+ O(mr^2/{\cal R}^3) \biggr], 
\label{3.3.13}
\end{eqnarray}
where $\gamma^{\rm tail}_{\alpha'\beta'}$ is the tensor of
Eq.~(\ref{3.3.12}) evaluated at $x'$, and
\begin{equation} 
\gamma^{\rm tail}_{\alpha'\beta'\gamma'}(x')  
= 4m \int_{-\infty}^{u^-} \nabla_{\gamma'}
G_{\alpha'\beta'\mu\nu}(x',z) u^\mu u^\nu\, d\tau 
\label{3.3.14}
\end{equation}
emerges during the computation of 
\[
\nabla_{\gamma'} \gamma^{\rm tail}_{\alpha'\beta'} = 4m 
(\partial_{\gamma'} u) V_{\alpha'\beta'\delta'\epsilon'} u^{\delta'}
u^{\epsilon'} + \gamma^{\rm tail}_{\alpha'\beta'\gamma'};
\] 
the term proportional to $\partial_{\gamma'} u$ disappears after
contraction with $\sigma^{\gamma'}$.    

At this stage we introduce the fields  
\begin{eqnarray}
h^{\rm tail}_{\alpha'\beta'} &=& 4m \int_{-\infty}^{u^-} 
\biggl( G_{\alpha'\beta'\mu\nu} 
- \frac{1}{2} g_{\alpha'\beta'} 
G^{\delta'}_{\ \delta'\mu\nu} \biggr) 
\nonumber \\ & & \hspace*{40pt} \mbox{} \times  
u^\mu u^\nu\, d\tau,   
\label{3.3.15} \\ 
h^{\rm tail}_{\alpha'\beta'\gamma'} &=& 4m \int_{-\infty}^{u^-}  
\nabla_{\gamma'} \biggl( G_{\alpha'\beta'\mu\nu} 
- \frac{1}{2} g_{\alpha'\beta'} 
G^{\delta'}_{\ \delta'\mu\nu} \biggr)
\nonumber \\ & & \hspace*{40pt} \mbox{} \times  
u^\mu u^\nu\, d\tau  
\label{3.3.16}
\end{eqnarray}
and recognize that the metric perturbation obtained from
Eqs.~(\ref{3.3.2}) and (\ref{3.3.13}) is 
\begin{eqnarray} 
h_{\alpha\beta}(x) &=& g^{\alpha'}_{\ \alpha} g^{\beta'}_{\ \beta}   
\biggl[ \frac{2m}{r} \Bigl( 2u_{\alpha'} u_{\beta'} 
+ g_{\alpha'\beta'} \Bigr) + h^{\rm tail}_{\alpha'\beta'}  
\nonumber \\ & & \mbox{} 
+ r h^{\rm tail}_{\alpha'\beta'\gamma'} \bigl( u^{\gamma'} +
\Omega^c \base{\gamma'}{c} \bigr) 
\nonumber \\ & & \mbox{} 
+ O(mr^2/{\cal R}^3) \biggr].
\label{3.3.17}
\end{eqnarray}
This is the desired expansion of the metric perturbation in powers of
$r/{\cal R}$. Our next task will be to calculate the components of
this tensor in the retarded coordinates $(u,r\Omega^a)$. 

The first step of this computation is to decompose $h_{\alpha\beta}$ 
in the tetrad $(\base{\alpha}{0},\base{\alpha}{a})$ that is obtained
by parallel transport of $(u^{\alpha'},\base{\alpha'}{a})$ on the null
geodesic that links $x$ to its corresponding retarded point $x' \equiv 
z(u)$ on the world line. (The vectors are parallel transported in the
background spacetime.) The projections are 
\begin{eqnarray}
h_{00}(u,r,\Omega^a) &\equiv& 
h_{\alpha\beta} \base{\alpha}{0} \base{\beta}{0} 
= \frac{2m}{r} + h^{\rm tail}_{00}(u) 
\nonumber \\ & & \mbox{} 
+ r \bigl[ h^{\rm tail}_{000}(u)
+ h^{\rm tail}_{00c}(u) \Omega^c \bigr] 
\nonumber \\ & & \mbox{} 
+ O(mr^2/{\cal R}^3), 
\label{3.3.18} \\ 
h_{0b}(u,r,\Omega^a) &\equiv& 
h_{\alpha\beta} \base{\alpha}{0} \base{\beta}{b} 
= h^{\rm tail}_{0b}(u) 
\nonumber \\ & & \mbox{} 
+ r \bigl[ h^{\rm tail}_{0b0}(u)
+ h^{\rm tail}_{0bc}(u) \Omega^c \bigr] 
\nonumber \\ & & \mbox{} 
+ O(mr^2/{\cal R}^3), 
\label{3.3.19} \\ 
h_{ab}(u,r,\Omega^a) &\equiv& 
h_{\alpha\beta} \base{\alpha}{a} \base{\beta}{b} 
= \frac{2m}{r}\delta_{ab} + h^{\rm tail}_{ab}(u) 
\nonumber \\ & & \mbox{} 
+ r \bigl[ h^{\rm tail}_{ab0}(u) 
+ h^{\rm tail}_{abc}(u) \Omega^c \bigr] 
\nonumber \\ & & \mbox{} 
+ O(mr^2/{\cal R}^3).
\label{3.3.20} 
\end{eqnarray} 
On the right-hand side we have the frame components of  
$h^{\rm tail}_{\alpha'\beta'}$ and 
$h^{\rm tail}_{\alpha'\beta'\gamma'}$ taken with respect 
to the tetrad $(u^{\alpha'},\base{\alpha'}{a})$; these 
are functions of retarded time $u$ only. 

The perturbation is now expressed as  
\[
h_{\alpha\beta} = h_{00} \base{0}{\alpha} \base{0}{\beta} 
+ h_{0b} \bigl( \base{0}{\alpha} \base{b}{\beta} 
+ \base{b}{\alpha} \base{0}{\beta} \bigr)   
+ h_{ab} \base{a}{\alpha} \base{b}{\beta} 
\]
and its components are obtained by involving Eqs.~(\ref{2.6.1}) and 
(\ref{2.6.2}), which list the components of the tetrad vectors in the  
retarded coordinates; this is the second (and longest) step of the 
computation. Noting that $a_a$ and $\omega_{ab}$ can both be set equal 
to zero in these equations (because they would produce negligible
terms of order $m^2$ in $h_{\alpha\beta}$), and that $S_{ab}$, $S_a$, 
and $S$ can all be expressed in terms of the tidal fields 
${\cal E}^*$, ${\cal E}^*_a$, ${\cal E}^*_{ab}$, ${\cal B}^*_a$, and  
${\cal B}^*_{ab}$ using Eqs.~(\ref{2.8.10})--(\ref{2.8.12}), we arrive 
at   
\begin{eqnarray} 
h_{uu} &=& \frac{2m}{r} + h^{\rm tail}_{00} + r \bigl( 2m {\cal E}^* 
+ h^{\rm tail}_{000} + h^{\rm tail}_{00a} \Omega^a \bigr) 
\nonumber \\ & & \mbox{} 
+ O(mr^2/{\cal R}^3), 
\label{3.3.21} \\ 
h_{ua} &=& \frac{2m}{r}\Omega_a + h^{\rm tail}_{0a} 
+ \Omega_a h^{\rm tail}_{00} + r \biggl[ 2m {\cal E}^* \Omega_a 
\nonumber \\ & & \mbox{} 
+ \frac{2m}{3} \bigl( {\cal E}^*_a + {\cal B}^*_a \bigr) 
+ h^{\rm tail}_{0a0} + \Omega_a h^{\rm tail}_{000} 
+ h^{\rm tail}_{0ab} \Omega^b 
\nonumber \\ & & \mbox{} 
+ \Omega_a h^{\rm tail}_{00b} \Omega^b \biggr] 
+ O(mr^2/{\cal R}^3),
\label{3.3.22} \\ 
h_{ab} &=& \frac{2m}{r} \bigl(\delta_{ab} + \Omega_a\Omega_b \bigr) 
+ \Omega_a \Omega_b h^{\rm tail}_{00} + \Omega_a h^{\rm tail}_{0b} 
+ \Omega_b h^{\rm tail}_{0a} 
\nonumber \\ & & \mbox{} 
+ h^{\rm tail}_{ab} 
+ r \biggl[ -\frac{2m}{3} \Bigl( {\cal E}^*_{ab} 
+ \Omega_a {\cal E}^*_b + {\cal E}^*_a \Omega_b + {\cal B}^*_{ab} 
\nonumber \\ & & \mbox{} 
+ \Omega_a {\cal B}^*_b + \Omega_b {\cal B}^*_a \Bigr)    
+ \Omega_a \Omega_b \bigl( h^{\rm tail}_{000} + h^{\rm tail}_{00c}
\Omega^c \bigr) 
\nonumber \\ & & \mbox{} 
+ \Omega_a \bigl( h^{\rm tail}_{0b0} 
+ h^{\rm tail}_{0bc} \Omega^c \bigr) + \Omega_b 
\bigl( h^{\rm tail}_{0a0} + h^{\rm tail}_{0ac} \Omega^c \bigr) 
\nonumber \\ & & \mbox{} 
+ \bigl( h^{\rm tail}_{ab0} + h^{\rm tail}_{abc} \Omega^c \bigr)
\biggr] + O(mr^2/{\cal R}^3).
\label{3.3.23}
\end{eqnarray} 
These are the {\it coordinate components} of the metric perturbation  
$h_{\alpha\beta}$ in the retarded coordinates $(u,r\Omega^a)$,
expressed in terms of {\it frame components} of the tail fields 
$h^{\rm tails}_{\alpha'\beta'}$ and 
$h^{\rm tails}_{\alpha'\beta'\gamma'}$. The perturbation is expanded  
in powers of $r/{\cal R}$ and it also involves the tidal gravitational
fields of the background spacetime. 

The external-zone metric is obtained by adding $g_{\alpha\beta}$ as 
given by Eqs.~(\ref{3.3.7})--(\ref{3.3.9}) to $h_{\alpha\beta}$ as 
given by Eqs.~(\ref{3.3.21})--(\ref{3.3.23}). The final result is 
\begin{eqnarray} 
{\sf g}_{uu} &=& -1 - r^2 {\cal E}^* + O(r^3/{\cal R}^3) 
\nonumber \\ & & \mbox{} 
+ \frac{2m}{r} + h^{\rm tail}_{00} + r \bigl( 2m {\cal E}^* 
- 2 a_a \Omega^a + h^{\rm tail}_{000} 
\nonumber \\ & & \mbox{} 
+ h^{\rm tail}_{00a} \Omega^a \bigr) + O(mr^2/{\cal R}^3),  
\label{3.3.24} \\ 
{\sf g}_{ua} &=& -\Omega_a + \frac{2}{3} r^2 \bigl(
{\cal E}^*_a  + {\cal B}^*_a \bigr) + O(r^3/{\cal R}^3)
\nonumber \\ & & \mbox{} 
+ \frac{2m}{r}\Omega_a + h^{\rm tail}_{0a} 
+ \Omega_a h^{\rm tail}_{00} + r \biggl[ 2m {\cal E}^* \Omega_a 
\nonumber \\ & & \mbox{} 
+ \frac{2m}{3} \bigl( {\cal E}^*_a + {\cal B}^*_a \bigr)    
+ \bigr(\delta_a^{\ b} - \Omega_a \Omega^b \bigr) a_b 
- \omega_{ab} \Omega^b 
\nonumber \\ & & \mbox{} 
+ h^{\rm tail}_{0a0} 
+ \Omega_a h^{\rm tail}_{000} + h^{\rm tail}_{0ab} \Omega^b 
+ \Omega_a h^{\rm tail}_{00b} \Omega^b \biggr] 
\nonumber \\ & & \mbox{} 
+ O(mr^2/{\cal R}^3),
\label{3.3.25} \\ 
{\sf g}_{ab} &=& \delta_{ab} - \Omega_a \Omega_b - \frac{1}{3} r^2   
\bigl( {\cal E}^*_{ab} + {\cal B}^*_{ab} \bigr) + O(r^3/{\cal R}^3) 
\nonumber \\ & & \mbox{}
+ \frac{2m}{r} \bigl(\delta_{ab} + \Omega_a\Omega_b \bigr) 
+ \Omega_a \Omega_b h^{\rm tail}_{00} + \Omega_a h^{\rm tail}_{0b} 
\nonumber \\ & & \mbox{} 
+ \Omega_b h^{\rm tail}_{0a} + h^{\rm tail}_{ab} 
+ r \biggl[ -\frac{2m}{3} \Bigl( {\cal E}^*_{ab} 
+ \Omega_a {\cal E}^*_b 
\nonumber \\ & & \mbox{} 
+ {\cal E}^*_a \Omega_b + {\cal B}^*_{ab} 
+ \Omega_a {\cal B}^*_b + \Omega_b {\cal B}^*_a \Bigr)    
+ \Omega_a \Omega_b \bigl( h^{\rm tail}_{000} 
\nonumber \\ & & \mbox{} 
+ h^{\rm tail}_{00c} \Omega^c \bigr) 
+ \Omega_a \bigl( h^{\rm tail}_{0b0} 
+ h^{\rm tail}_{0bc} \Omega^c \bigr) + \Omega_b 
\bigl( h^{\rm tail}_{0a0} 
\nonumber \\ & & \mbox{} 
+ h^{\rm tail}_{0ac} \Omega^c \bigr) 
+ \bigl( h^{\rm tail}_{ab0} + h^{\rm tail}_{abc} \Omega^c \bigr)
\biggr] 
\nonumber \\ & & \mbox{} 
+ O(mr^2/{\cal R}^3).
\label{3.3.26}  
\end{eqnarray} 

\subsection{Matching: motion of the black hole in the background
spacetime}  

Comparison of Eqs.~(\ref{3.2.9})--(\ref{3.2.11}) and
Eqs.~(\ref{3.3.24})--(\ref{3.3.26}) reveals that the internal-zone  
and external-zone metrics do no match in the buffer zone. But as the 
metrics are expressed in different coordinate systems, this
mismatch is hardly surprising. A meaningful comparison of the two
metrics must therefore come after a transformation from the external 
coordinates $(u,r\Omega^a)$ to the internal coordinates
$(\bar{u},\bar{r}\bar{\Omega}^a)$. This transformation is worked out
in Appendix C, and it puts the external-zone metric in its final form  
\begin{eqnarray} 
{\sf g}_{\bar{u}\bar{u}} &=& - 1 - \bar{r}^2 \bar{\cal E}^* 
+ O(\bar{r}^3/{\cal R}^3) 
\nonumber \\ & & \mbox{} 
+ \frac{2m}{\bar{r}} + \bar{r} \biggl[ 4m \bar{\cal E}^* 
- 2 \biggl( a_a -\frac{1}{2} h^{\rm tail}_{00a} 
+ h^{\rm tail}_{0a0} \biggr) \bar{\Omega}^a \biggr] 
\nonumber \\ & & \mbox{} 
+ O(m \bar{r}^2/{\cal R}^3), 
\label{3.4.1} \\ 
{\sf g}_{\bar{u}\bar{a}} &=& -\bar{\Omega}_a + \frac{2}{3} \bar{r}^2
\bigl( \bar{\cal E}^*_a + \bar{\cal B}^*_a \bigr) 
+ O(\bar{r}^3/{\cal R}^3) 
\nonumber \\ & & \mbox{} 
+ \bar{r} \biggl[ -\frac{4m}{3} \bigl( \bar{\cal E}^*_a 
+ \bar{\cal B}^*_a \bigr) + (\delta_a^{\ b} - \bar{\Omega}_a
\bar{\Omega}^b \bigr) \biggl( a_b 
\nonumber \\ & & \mbox{} 
-\frac{1}{2} h^{\rm tail}_{00b} + h^{\rm tail}_{0b0} \biggr) 
- \bigl( \omega_{ab} 
- h^{\rm tail}_{0[ab]} \bigr) \bar{\Omega}^b \biggr] 
\nonumber \\ & & \mbox{} 
+ O(m \bar{r}^2/{\cal R}^3), 
\label{3.4.2} \\ 
{\sf g}_{\bar{a}\bar{b}} &=& \delta_{ab} - \bar{\Omega}_a
\bar{\Omega}_b - \frac{1}{3} \bar{r}^2 \bigl( \bar{\cal E}^*_{ab}  
+ \bar{\cal B}^*_{ab} \bigr) 
\nonumber \\ & & \mbox{} 
+ O(\bar{r}^3/{\cal R}^3) + O(m \bar{r}^2/{\cal R}^3). 
\label{3.4.3} 
\end{eqnarray} 
Except for the terms involving $a_a$ and $\omega_{ab}$, this metric
is equal to ${\sf g}(\mbox{internal zone})$ as given by
Eqs.~(\ref{3.2.9})--(\ref{3.2.11}) linearized with respect to 
$m$.    

A precise match between ${\sf g}(\mbox{external zone})$ and 
${\sf g}(\mbox{internal zone})$ is produced when we impose the
relations 
\begin{equation} 
a_a = \frac{1}{2} h^{\rm tail}_{00a} - h^{\rm tail}_{0a0}
\label{3.4.4}
\end{equation}
and 
\begin{equation}
\omega_{ab} = h^{\rm tail}_{0[ab]}. 
\label{3.4.5}
\end{equation} 
While Eq.~(\ref{3.4.4}) tells us how the black hole moves in the
background spacetime, Eq.~(\ref{3.4.5}) indicates that the vectors
$\base{\mu}{a}$ are not Fermi-Walker transported on the world line.   

The black hole's acceleration vector $a^\mu = a^a \base{\mu}{a}$ can
be constructed from the frame components listed in
Eq.~(\ref{3.4.4}). A straightforward computation gives 
\begin{equation}
a^\mu = -\frac{1}{2} \bigl( g^{\mu\nu} + u^\mu
u^\nu \bigr) \bigl( 2 h^{\rm tail}_{\nu\lambda\rho} 
- h^{\rm tail}_{\lambda\rho\nu} \bigr) u^\lambda u^\rho,  
\label{3.4.6} 
\end{equation}   
where the tail integral 
\begin{eqnarray}
h^{\rm tail}_{\mu\nu\lambda} &=& 4 m \int_{-\infty}^{\tau^-} 
\nabla_\lambda \biggl( G_{\mu\nu\mu'\nu'}
- \frac{1}{2} g_{\mu\nu} G^{\rho}_{\ \rho\mu'\nu'}
\biggr)(\tau,\tau') 
\nonumber \\ & & \hspace*{40pt} \mbox{} \times 
u^{\mu'} u^{\nu'}\, d\tau' 
\label{3.4.7} 
\end{eqnarray}     
was previously defined by Eq.~(\ref{3.3.16}). Here, the unprimed
indices refer to the current position $z(\tau)$ on the world line,
while the primed indices refer to a prior position $z(\tau')$; the
integral is cut short at $\tau' = \tau^-$ in the manner defined by
Eq.~(\ref{3.3.12}). These are the MiSaTaQuWa equations of motion, as
they were first presented by Mino, Sasaki, and Tanaka
\cite{mino-etal:97}, and later rederived by Quinn and Wald
\cite{quinn-wald:97}.    

Substituting Eqs.~(\ref{3.4.4}) and (\ref{3.4.5}) into
Eq.~(\ref{2.1.1}) gives the following transport equation for the
tetrad vectors: 
\begin{eqnarray} 
\frac{D \base{\mu}{a}}{d\tau} &=& -\frac{1}{2} u^\mu      
\bigl( 2 h^{\rm tail}_{\nu\lambda\rho} 
- h^{\rm tail}_{\nu\rho\lambda} \bigr) u^\nu \base{\lambda}{a} u^\rho 
\nonumber \\ & & \mbox{} 
+ \bigl(g^{\mu\rho} + u^\mu u^\rho \bigr) 
h^{\rm tail}_{\nu[\lambda\rho]} u^\nu \base{\lambda}{a}. 
\label{3.4.8}
\end{eqnarray} 
This can also be written in the alternative form 
\begin{equation} 
\frac{D \base{\mu}{a}}{d\tau} = -\frac{1}{2} \Bigl( 
u^\mu \base{\lambda}{a} u^\rho + g^{\mu\lambda} \base{\rho}{a} 
- g^{\mu\rho} \base{\lambda}{a} \Bigr) u^\nu 
h^{\rm tail}_{\nu\lambda\rho} 
\label{3.4.9}
\end{equation} 
that was first proposed by Mino, Sasaki, and Tanaka. Both equations 
state that in the background spacetime, the tetrad vectors are not
Fermi-Walker transported on $\gamma$; the rotation tensor is nonzero
and given by Eq.~(\ref{3.4.5}).   

\begin{acknowledgments} 
This work was supported by the Natural Sciences and Engineering
Research Council of Canada. Conversations with Claude Barrab\`es and
Werner Israel were instrumental during the early stage of my work on
the retarded coordinates. Conversations with Steve Detweiler helped me 
clarify my thoughts on matched asymptotic expansions.   
\end{acknowledgments} 

\appendix

\section{Spherical harmonics}

In this Appendix I provide explicit expressions for the scalar,
vectorial, and tensorial spherical harmonics introduced in Sec.~II H.  
All harmonics are of degree $l=2$. 

The scalar harmonics are
\begin{eqnarray*} 
Y^0 &=& \frac{1}{2} (3\cos^2\theta - 1), \\ 
Y^{1c} &=& \sin\theta \cos\theta \cos\phi, \\ 
Y^{1s} &=& \sin\theta \cos\theta \sin\phi, \\ 
Y^{2c} &=& \sin^2\theta \cos 2\phi, \\ 
Y^{2s} &=& \sin^2\theta \sin 2\phi.
\end{eqnarray*} 
The vectorial harmonics are defined by $Y^{\sf m}_A = Y^{\sf m}_{:A}$
(even parity) and $X^{\sf m}_A = -\varepsilon_A^{\ B} Y^{\sf m}_{:B}$
(odd parity), where a colon indicates covariant differentiation with
respect to a connection compatible with $\Omega_{AB} = 
\mbox{diag}(1,\sin^2\theta)$, and $\varepsilon_{AB}$ is the
two-dimensional Levi-Civita tensor. Explicitly, 
\begin{eqnarray*} 
Y^0_\theta &=& -3\sin\theta\cos\theta, \\
Y^0_\phi &=& 0, \\ 
Y^{1c}_\theta &=& (2\cos^2\theta-1)\cos\phi, \\ 
Y^{1c}_\phi &=& -\sin\theta\cos\theta \sin\phi, \\ 
Y^{1s}_\theta &=& (2\cos^2\theta-1)\sin\phi, \\ 
Y^{1s}_\phi &=& \sin\theta\cos\theta \cos\phi, \\
Y^{2c}_\theta &=& 2\sin\theta\cos\theta\cos 2\phi, \\ 
Y^{2c}_\phi &=& -2\sin^2\theta\sin 2\phi, \\ 
Y^{2s}_\theta &=& 2\sin\theta\cos\theta\sin 2\phi, \\ 
Y^{2c}_\phi &=& 2\sin^2\theta\cos 2\phi,
\end{eqnarray*} 
and 
\begin{eqnarray*} 
X^0_\theta &=& 0, \\ 
X^0_\phi &=& -3\sin^2\theta\cos\theta, \\ 
X^{1c}_\theta &=& \cos\theta\sin\phi, \\
X^{1c}_\phi &=& (2\cos^2\theta-1)\sin\theta\cos\phi, \\ 
X^{1s}_\theta &=& -\cos\theta\cos\phi, \\
X^{1s}_\phi &=& (2\cos^2\theta-1)\sin\theta\sin\phi, \\ 
X^{2c}_\theta &=& 2\sin\theta\sin 2\phi, \\
X^{2c}_\phi &=& 2\sin^2\theta\cos\theta\cos 2\phi, \\ 
X^{2s}_\theta &=& -2\sin\theta\cos 2\phi, \\
X^{2s}_\phi &=& 2\sin^2\theta\cos\theta\sin 2\phi.
\end{eqnarray*} 
The tensorial harmonics are defined by $Y^{\sf m}_{AB} = 
Y^{\sf m}_{:AB}$ (even parity) and $X^{\sf m}_{AB} = 
-X^{\sf m}_{(A:B)}$ (odd parity). Explicitly,  
\begin{eqnarray*} 
Y^0_{\theta\theta} &=& -3(2\cos^2\theta-1), \\ 
Y^0_{\theta\phi} &=& 0, \\ 
Y^0_{\phi\phi} &=& -3\sin^2\theta\cos^2\theta, \\ 
Y^{1c}_{\theta\theta} &=& -4\sin\theta\cos\theta\cos\phi, \\ 
Y^{1c}_{\theta\phi} &=& \sin^2\theta\sin\phi, \\ 
Y^{1c}_{\phi\phi} &=& -2\sin^3\theta\cos\theta\cos\phi, \\ 
Y^{1s}_{\theta\theta} &=& -4\sin\theta\cos\theta\sin\phi, \\ 
Y^{1s}_{\theta\phi} &=& -\sin^2\theta\cos\phi, \\ 
Y^{1s}_{\phi\phi} &=& -2\sin^3\theta\cos\theta\sin\phi, \\ 
Y^{2c}_{\theta\theta} &=& 2(2\cos^2\theta-1)\cos 2\phi, \\ 
Y^{2c}_{\theta\phi} &=& -2\sin\theta\cos\theta\sin 2\phi, \\ 
Y^{2c}_{\phi\phi} &=& 2\sin^2\theta(\cos^2\theta-2)\cos 2\phi, \\ 
Y^{2s}_{\theta\theta} &=& 2(2\cos^2\theta-1)\sin 2\phi, \\ 
Y^{2s}_{\theta\phi} &=& 2\sin\theta\cos\theta\cos 2\phi, \\ 
Y^{2s}_{\phi\phi} &=& 2\sin^2\theta(\cos^2\theta-2)\sin 2\phi,
\end{eqnarray*}
and 
\begin{eqnarray*} 
X^0_{\theta\theta} &=& 0, \\ 
X^0_{\theta\phi} &=& -\frac{3}{2} \sin^3\theta, \\ 
X^0_{\phi\phi} &=& 0, \\ 
X^{1c}_{\theta\theta} &=& \sin\theta\sin\phi, \\ 
X^{1c}_{\theta\phi} &=& \sin^2\theta\cos\theta\cos\phi, \\ 
X^{1c}_{\phi\phi} &=& -\sin^3\theta\sin\phi, \\ 
X^{1s}_{\theta\theta} &=& -\sin\theta\cos\phi, \\ 
X^{1s}_{\theta\phi} &=& \sin^2\theta\cos\theta\sin\phi, \\ 
X^{1s}_{\phi\phi} &=& \sin^3\theta\cos\phi, \\ 
X^{2c}_{\theta\theta} &=& -2\cos\theta\sin 2\phi, \\ 
X^{2c}_{\theta\phi} &=& -\sin\theta(\cos^2\theta+1)\cos 2\phi, \\ 
X^{2c}_{\phi\phi} &=& 2\sin^2\theta\cos\theta\sin 2\phi, \\ 
X^{2s}_{\theta\theta} &=& 2\cos\theta\cos 2\phi, \\ 
X^{2s}_{\theta\phi} &=& -\sin\theta(\cos^2\theta+1)\sin 2\phi, \\ 
X^{2c}_{\phi\phi} &=& -2\sin^2\theta\cos\theta\cos 2\phi.
\end{eqnarray*}  

\section{Calculation of the metric perturbations} 

In this Appendix I derive the form of the functions $e_1$, $e_2$,
$e_3$, $b_2$, and $b_3$ that appear in Sec.~III B. For this it is
sufficient to take, say, ${\cal E}_{12} = {\cal E}_{21}$ and ${\cal
B}_{12} = {\cal B}_{21}$ as the only nonvanishing components of the
tidal fields ${\cal E}_{ab}$ and ${\cal B}_{ab}$. And since the
equations for even-parity and odd-parity perturbations decouple
\cite{regge-wheeler:57, zerilli:70}, each case can be considered
separately.  

Including only even-parity perturbations,
Eqs.~(\ref{3.2.3})--(\ref{3.2.6}) become   
\begin{eqnarray*} 
{\sf g}_{\bar{u}\bar{u}} &=& - f \bigr( 1 + \bar{r}^2 e_1   
{\cal E}_{12} \sin^2\bar{\theta} \sin2\bar\phi \bigr), \\
{\sf g}_{\bar{u}\bar{r}} &=& - 1, \\ 
{\sf g}_{\bar{u}\bar{\theta}} &=& \frac{2}{3} \bar{r}^3 e_2   
{\cal E}_{12} \sin\bar{\theta} \cos\bar{\theta} \sin2\bar{\phi}, \\
{\sf g}_{\bar{u}\bar{\phi}} &=& \frac{2}{3} \bar{r}^3 e_2   
{\cal E}_{12} \sin^2\bar{\theta} \cos2\bar{\phi}, \\
{\sf g}_{\bar{\theta}\bar{\theta}} &=& \bar{r}^2 
- \frac{1}{3} \bar{r}^4 e_3 {\cal E}_{12} 
(1 + \cos^2\bar{\theta}) \sin2\bar{\phi}, \\ 
{\sf g}_{\bar{\theta}\bar{\phi}} &=& - \frac{2}{3} \bar{r}^4       
e_3 {\cal E}_{12} \sin\bar{\theta}\cos\bar{\theta} \cos2\bar{\phi}, \\
{\sf g}_{\bar{\phi}\bar{\phi}} &=& \bar{r}^2\sin^2\bar{\theta} +
\frac{1}{3} \bar{r}^4 e_3 {\cal E}_{12} \sin^2\bar{\theta} (1 +
\cos^2\bar{\theta}) \sin2\bar{\phi}. 
\end{eqnarray*}
This metric is then substituted into the vacuum Einstein field
equations. Computing the Einstein tensor is simplified by
linearizing with respect to ${\cal E}_{12}$ and discarding its
derivatives with respect to $\bar{u}$: Since the time scale over which
${\cal E}_{ab}$ changes is of order ${\cal R}$, the ratio between
temporal and spatial derivatives is of order $\bar{r}/{\cal R}$ and
therefore small in the internal zone; the temporal derivatives can be
consistently neglected. The field equations produce ordinary
differential equations to be satisfied by the functions $e_1$, $e_2$,
and $e_3$. Those are easily decoupled, and demanding that the
functions all approach unity as $r \to \infty$ and be well-behaved at
$r=2m$ yields the unique solutions 
\[
e_1(\bar{r}) = e_2(\bar{r}) = f, \qquad  
e_3(\bar{r}) = 1 - \frac{2m^2}{\bar{r}^2}, 
\]
as was stated in Eq.~(\ref{3.2.7}).  

Switching now to odd-parity perturbations,  
Eqs.~(\ref{3.2.3})--(\ref{3.2.6}) become  
\begin{eqnarray*} 
{\sf g}_{\bar{u}\bar{u}} &=& - f, \\  
{\sf g}_{\bar{u}\bar{r}} &=& - 1, \\  
{\sf g}_{\bar{u}\bar{\theta}} &=& -\frac{2}{3} \bar{r}^3 b_2   
{\cal B}_{12} \sin\bar{\theta} \cos2\bar{\phi}, \\  
{\sf g}_{\bar{u}\bar{\phi}} &=& \frac{2}{3} \bar{r}^3 b_2   
{\cal B}_{12} \sin^2\bar{\theta} \cos\bar{\theta} \sin2\bar{\phi}, \\   
{\sf g}_{\bar{\theta}\bar{\theta}} &=& \bar{r}^2 
+ \frac{2}{3} \bar{r}^4 b_3 {\cal B}_{12} \cos\bar{\theta}
\cos2\bar{\phi}, \\  
{\sf g}_{\bar{\theta}\bar{\phi}} &=& - \frac{1}{3} \bar{r}^4      
b_3 {\cal B}_{12} \sin\bar{\theta} (1 + \cos^2\bar{\theta})
\sin2\bar{\phi}, \\ 
{\sf g}_{\bar{\phi}\bar{\phi}} &=& \bar{r}^2\sin^2\bar{\theta} -
\frac{2}{3} \bar{r}^4 b_3 {\cal B}_{12} \sin^2\bar{\theta}
\cos\bar{\theta} \cos2\bar{\phi}.  
\end{eqnarray*} 
Following the same procedure, we arrive at 
\[
b_2(\bar{r}) = f, \qquad  
b_3(\bar{r}) = 1, 
\]
as was stated in Eq.~(\ref{3.2.8}). 

\section{Transformation from external to internal coordinates} 

Our task in this Appendix is to construct the transformation from the
external coordinates $(u,r\Omega^a)$ to the internal coordinates
$(\bar{u},\bar{r}\bar{\Omega}^a)$. We shall proceed in three
stages. The first stage of the transformation,
$(u,r\Omega^a) \to (u',r'\Omega^{\prime a})$, will be seen to remove 
unwanted terms of order $m/r$ in ${\sf g}_{\alpha\beta}$, as listed in
Eqs.~(\ref{3.3.24})--(\ref{3.3.26}). The second stage, 
$(u',r'\Omega^{\prime a}) \to (u'',r''\Omega^{\prime\prime a})$, will
remove all terms of order $m/{\cal R}$ in 
${\sf g}_{\alpha'\beta'}$. Finally, the third stage 
$(u'',r''\Omega^{\prime\prime a}) \to (\bar{u},\bar{r}\bar{\Omega}^a)$ 
will produce the desired internal coordinates and return the metric
in the form of Eqs.~(\ref{3.4.1})--(\ref{3.4.3}).  

\begin{widetext} 
The first stage of the coordinate transformation is 
\begin{eqnarray*} 
u' &=& u - 2m \ln r, \\ 
x^{\prime a} &=& \Bigl( 1 + \frac{m}{r} \Bigr) x^a,   
\end{eqnarray*} 
and it affects the metric at orders $m/r$ and $mr/{\cal R}^2$. This  
transformation redefines the radial coordinate --- $r \to r' = r + m$
--- and incorporates in $u'$ the gravitational time delay contributed
by the small mass $m$. After performing the coordinate transformation
the metric becomes    
\begin{eqnarray*} 
{\sf g}_{u'u'} &=& -1 - r^{\prime 2} {\cal E}^{\prime *} 
+ O(r^{\prime 3}/{\cal R}^3)  
\nonumber \\ & & \mbox{} 
+ \frac{2m}{r'} + h^{\rm tail}_{00} + r' \bigl( 4m {\cal E}^{\prime *}  
- 2 a_a \Omega^{\prime a} + h^{\rm tail}_{000} + h^{\rm tail}_{00a}
\Omega^{\prime a} \bigr) + O(mr^{\prime 2}/{\cal R}^3), \\ 
{\sf g}_{u'a'} &=& -\Omega'_a + \frac{2}{3} r^{\prime 2} \bigl(
{\cal E}^{\prime *}_a  + {\cal B}^{\prime *}_a \bigr) 
+ O(r^{\prime 3}/{\cal R}^3) 
\nonumber \\ & & \mbox{} 
+ h^{\rm tail}_{0a} 
+ \Omega'_a h^{\rm tail}_{00} + r' \biggl[ - \frac{4m}{3} 
\bigl( {\cal E}^{\prime *}_a + {\cal B}^{\prime *}_a \bigr)     
+ \bigr(\delta_a^{\ b} - \Omega'_a \Omega^{\prime b} \bigr) a_b  
\nonumber \\ & & \mbox{} 
- \omega_{ab} \Omega^{\prime b} + h^{\rm tail}_{0a0}  
+ \Omega'_a h^{\rm tail}_{000} + h^{\rm tail}_{0ab} \Omega^{\prime b}  
+ \Omega'_a h^{\rm tail}_{00b} \Omega^{\prime b} \biggr]  
+ O(mr^{\prime 2}/{\cal R}^3), \\ 
{\sf g}_{a'b'} &=& \delta_{ab} - \Omega'_a \Omega'_b 
- \frac{1}{3} r^{\prime 2} \bigl( {\cal E}^{\prime *}_{ab} 
+ {\cal B}^{\prime *}_{ab} \bigr) + O(r^{\prime 3}/{\cal R}^3)  
\nonumber \\ & & \mbox{}
+ \Omega'_a \Omega'_b h^{\rm tail}_{00} + \Omega'_a h^{\rm tail}_{0b}  
+ \Omega'_b h^{\rm tail}_{0a} + h^{\rm tail}_{ab} 
+ r' \biggl[ \frac{2m}{3} \Bigl( {\cal E}^{\prime *}_{ab} 
+ \Omega'_a {\cal E}^{\prime *}_b 
+ {\cal E}^{\prime *}_a \Omega'_b 
\nonumber \\ & & \mbox{} 
+ {\cal B}^{\prime *}_{ab} 
+ \Omega'_a {\cal B}^{\prime *}_b 
+ \Omega'_b {\cal B}^{\prime *}_a \Bigr)     
+ \Omega'_a \Omega'_b \bigl( h^{\rm tail}_{000} + h^{\rm tail}_{00c}
\Omega^{\prime c} \bigr)  
+ \Omega'_a \bigl( h^{\rm tail}_{0b0} 
+ h^{\rm tail}_{0bc} \Omega^{\prime c} \bigr) 
\nonumber \\ & & \mbox{} 
+ \Omega'_b \bigl( h^{\rm tail}_{0a0} 
+ h^{\rm tail}_{0ac} \Omega^{\prime c} \bigr) 
+ \bigl( h^{\rm tail}_{ab0} 
+ h^{\rm tail}_{abc} \Omega^{\prime c} \bigr) \biggr] 
+ O(mr^{\prime 2}/{\cal R}^3).
\end{eqnarray*} 
This metric matches ${\sf g}(\mbox{internal zone})$ at orders $1$, 
$r^{\prime 2}/{\cal R}^2$, and $m/r'$, but there is still a mismatch
at orders $m/{\cal R}$ and $mr'/{\cal R}^2$. 

The second stage of the coordinate transformation is  
\begin{eqnarray*}
u'' &=& u' - \frac{1}{2} \int^{u'} h^{\rm tail}_{00}(u')\, du' 
- \frac{1}{2} r' \Bigl[ h^{\rm tail}_{00}(u') 
+ 2 h^{\rm tail}_{0a}(u') \Omega^{\prime a} 
+ h^{\rm tail}_{ab}(u') \Omega^{\prime a} \Omega^{\prime b} 
\Bigr], \\ 
x''_a &=& x'_a 
+ \frac{1}{2} h^{\rm tail}_{ab}(u') x^{\prime b},  
\end{eqnarray*}  
and it affects the metric at orders $m/{\cal R}$ and $mr/{\cal R}^2$. 
After performing this transformation the metric becomes   
\begin{eqnarray*} 
{\sf g}_{u''u''} &=& -1 
- r^{\prime\prime 2} {\cal E}^{\prime\prime *}  
+ O(r^{\prime\prime 3}/{\cal R}^3)  
\nonumber \\ & & \mbox{} 
+ \frac{2m}{r''} + r'' \biggl[ 4m {\cal E}^{\prime\prime *}  
- 2 \biggl(a_a -\frac{1}{2} h^{\rm tail}_{00a} + h^{\rm tail}_{0a0}
\biggr) \Omega^{\prime\prime a} \biggr] 
+ O(mr^{\prime\prime 2}/{\cal R}^3), \\ 
{\sf g}_{u''a''} &=& -\Omega''_a 
+ \frac{2}{3} r^{\prime\prime 2} \bigl(
{\cal E}^{\prime\prime *}_a  + {\cal B}^{\prime\prime *}_a \bigr)  
+ O(r^{\prime\prime 3}/{\cal R}^3) 
\nonumber \\ & & \mbox{} 
+ r'' \biggl[ - \frac{4m}{3} 
\bigl( {\cal E}^{\prime\prime *}_a + {\cal B}^{\prime\prime *}_a
\bigr) - 2 m {\cal E}_{ab} \Omega^{\prime\prime b} 
+ \bigr(\delta_a^{\ b} - \Omega''_a \Omega^{\prime\prime b}
\bigr) \biggl( a_b -\frac{1}{2} h^{\rm tail}_{00b} 
+ h^{\rm tail}_{0b0} \biggr) - \omega_{ab} \Omega^{\prime\prime b} 
\nonumber \\ & & \mbox{} 
+ \frac{1}{2} \Omega''_a h^{\rm tail}_{000} 
- \frac{1}{2} h^{\rm tail}_{ab0} \Omega^{\prime\prime b} 
+ h^{\rm tail}_{0ab} \Omega^{\prime\prime b}  
+ \frac{1}{2} \bigr(\delta_a^{\ b} 
+ \Omega''_a \Omega^{\prime\prime b} \bigr) h^{\rm tail}_{00b} \biggr]   
+ O(mr^{\prime\prime 2}/{\cal R}^3), \\ 
{\sf g}_{a''b''} &=& \delta_{ab} 
- \Omega''_a \Omega''_b  
- \frac{1}{3} r^{\prime\prime 2} \bigr( {\cal E}^{\prime\prime *}_{ab}  
+ {\cal B}^{\prime\prime *}_{ab} \bigr) 
+ O(r^{\prime\prime 3}/{\cal R}^3)   
\nonumber \\ & & \mbox{}
+ r'' \biggl[ \frac{2m}{3} \Bigl( {\cal E}^{\prime\prime *}_{ab} 
+ \Omega''_a {\cal E}^{\prime\prime *}_b 
+ {\cal E}^{\prime\prime *}_a \Omega''_b 
+ {\cal B}^{\prime\prime *}_{ab} 
+ \Omega''_a {\cal B}^{\prime\prime *}_b 
+ \Omega''_b {\cal B}^{\prime\prime *}_a \Bigr)     
+ \Omega''_a \Omega''_b \bigl( h^{\rm tail}_{000} + h^{\rm tail}_{00c}
\Omega^{\prime\prime c} \bigr)  
\nonumber \\ & & \mbox{} 
+ \Omega''_a \bigl( h^{\rm tail}_{0b0} 
+ h^{\rm tail}_{0bc} \Omega^{\prime\prime c} \bigr) 
+ \Omega''_b \bigl( h^{\rm tail}_{0a0} 
+ h^{\rm tail}_{0ac} \Omega^{\prime\prime c} \bigr) 
+ \bigl( h^{\rm tail}_{ab0} 
+ h^{\rm tail}_{abc} \Omega^{\prime\prime c} \bigr) \biggr] 
+ O(mr^{\prime\prime 2}/{\cal R}^3). 
\end{eqnarray*} 
To arrive at these expressions we had to involve the relations  
\begin{eqnarray*}  
\frac{d}{du''} h^{\rm tail}_{00} &=& h^{\rm tail}_{000}, \\
\frac{d}{du''} h^{\rm tail}_{0a} &=& h^{\rm tail}_{0a0}, \\
\frac{d}{du''} h^{\rm tail}_{ab} &=& 4m {\cal E}_{ab} 
+ h^{\rm tail}_{ab0}, 
\end{eqnarray*}
which are obtained by covariant differentiation of  
Eq.~(\ref{3.3.15}) in the direction of $u^{\alpha'}$. The metric now
matches ${\sf g}(\mbox{internal zone})$ at orders $1$, 
$r^{\prime\prime 2}/{\cal R}^2$, $m/r''$, and $m/{\cal R}$, but there 
is still a mismatch at order $mr''/{\cal R}^2$.  

The third and final stage of the coordinate transformation is 
\begin{eqnarray*} 
\bar{u} &=& u'' - \frac{1}{4} r^{\prime\prime 2} \Bigl[ 
h^{\rm tail}_{000} + \bigl(h^{\rm tail}_{00a} 
+ 2 h^{\rm tail}_{0a0} \bigr) \Omega^{\prime\prime a} 
+ \bigl( h^{\rm tail}_{ab0} + 2 h^{\rm tail}_{0ab} \bigr)    
\Omega^{\prime\prime a} \Omega^{\prime\prime b} 
+ h^{\rm tail}_{abc} \Omega^{\prime\prime a} \Omega^{\prime\prime b}
\Omega^{\prime\prime c} \Bigr], \\ 
\bar{x}_a &=& \biggl( 1 + \frac{m}{3} r'' {\cal E}_{bc}   
\Omega^{\prime\prime b} \Omega^{\prime\prime c} \biggr)
x''_a + \frac{1}{2} r^{\prime\prime 2} \biggl[ 
-\frac{1}{2} h^{\rm tail}_{00a} + h^{\rm tail}_{0a0} 
+ \biggl( h^{\rm tail}_{0ab} - h^{\rm tail}_{0ba} 
+ h^{\rm tail}_{ab0} + \frac{4m}{3} {\cal E}_{ab} \biggr)
\Omega^{\prime\prime b} 
\nonumber \\ & & \mbox{} 
+ \bigl(Q_{abc} - Q_{bca} + Q_{cab} \bigr) 
\Omega^{\prime\prime b} \Omega^{\prime\prime c} \biggr], 
\end{eqnarray*} 
where 
\[ 
Q_{abc} = \frac{1}{2} h^{\rm tail}_{abc} + \frac{m}{3}
\Bigl( \varepsilon_{acd} {\cal B}^d_{\ b} +
\varepsilon_{bcd} {\cal B}^d_{\ a} \Bigr).  
\]
This produces the metric of Eqs.~(\ref{3.4.1})--(\ref{3.4.3}). 
\end{widetext} 

\bibliography{motion}
\end{document}